\documentclass[journal]{IEEEtran}
\ifCLASSINFOpdf
\else
\fi

\usepackage{booktabs}
\usepackage{multirow}
\usepackage[utf8]{inputenc}
\usepackage{amsmath}
\usepackage{amssymb}
\usepackage{array}
\usepackage{mathtools}
\usepackage{braket}
\usepackage{graphicx}
\graphicspath{{./figures/}}
\usepackage{amsfonts}
\usepackage{algorithm}
\usepackage{algpseudocode}
\let\oldnl\nl% Store \nl in \oldnl
\newcommand{\nonl}{\renewcommand{\nl}{\let\nl\oldnl}}% Remove line number for one line

\usepackage{csquotes}
\usepackage{hhline}
\usepackage{amssymb}
\usepackage{listings}
\usepackage{color}
\definecolor{codegreen}{rgb}{0,0.6,0}
\definecolor{codegray}{rgb}{0.5,0.5,0.5}
\definecolor{codepurple}{rgb}{0.58,0,0.82}
\definecolor{backcolour}{rgb}{0.95,0.95,0.92}

\usepackage{subcaption}
\usepackage{titlesec}

\usepackage{dcolumn}
\usepackage{tabularx}
\usepackage{hyperref}
\hypersetup{
    colorlinks=true,
    linktoc=all,
    linkcolor=black,
    citecolor=magenta,
}
\usepackage{longtable}
\usepackage{braket}
\newcolumntype{C}{>{\centering\arraybackslash}X}
\usepackage[font=small]{caption}
\usepackage{tikz,xcolor,hyperref}% Make Orcid icon
\definecolor{lime}{HTML}{A6CE39}
\DeclareRobustCommand{\orcidicon}{%
    \begin{tikzpicture}
    \draw[lime, fill=lime] (0,0) 
    circle [radius=0.16] 
    node[white] {{\fontfamily{qag}\selectfont \tiny ID}};    \draw[white, fill=white] (-0.0625,0.095) 
    circle [radius=0.007];    \end{tikzpicture}
    \hspace{-2mm}}
\foreach \x in {A, ..., Z}{%
    \expandafter\xdef\csname orcid\x\endcsname{\noexpand\href{https://orcid.org/\csname orcidauthor\x\endcsname}{\noexpand\orcidicon}}
    }

\begin{document}
% correct bad hyphenation here
\hyphenation{op-tical net-works semi-conduc-tor}

\title{QTN-RQNN: Quantum Tensor Network-Based Recurrent Quantum Neural Network}
\author{Amit S. Patel\thanks{A. S. Patel is with the Department of Mechanical engineering, Dharmsinh Desai University, Nadiad, Gujarat, India. e-mail: (aspatel.mh@ddu.ac.in).}, Himanshu R. Patel\thanks{H. R. Patel is with the Department of Instrumentation \& Control Engineering, Dharmsinh Desai University, Nadiad, Gujarat, India. e-mail: (himanshupatel.ic@ddu.ac.in).}, Bikash~K.~Behera\thanks{B.~K. Behera is with the Bikash's Quantum (OPC) Pvt. Ltd., Mohanpur, WB, 741246, India and Università degli Studi di Cagliari, Via Is Mirrions, Cagliari, 09123, Italy, e-mail: (bikas.riki@gmail.com).}, Pranav Mehta\thanks{P. Mehta is with the Department of Mechanical Engineering, Dharmsinh Desai University, Nadiad-387001, Gujarat, India. e-mail: (Pranavmehta.mech@ddu.ac.in).}, Ghanshyam Tejani\thanks{G. Tejani is with the Department of Industrial Engineering and Management, Yuan Ze University, Taoyuan, 320315, Taiwan. e-mail: (gtejani@saturn.yzu.edu.tw).} and Deepak Parashar\thanks{D. Parashar is with the Manipal Institute of Technology, Manipal Academy of Higher Education, Manipal, India. e-mail: (deepak.parashar@manipal.edu).}.}
%<-this % stops a space

%
%
%
%\thanks{Manuscript received \today.}}


\maketitle
\begin{abstract}
Resource constraints and imperfect quantum operations remain major
challenges for quantum machine learning (QML) on noisy
intermediate-scale quantum (NISQ) devices, particularly for
fault-diagnosis tasks involving imbalanced and heterogeneous data.
This work proposes \textbf{QTN-RQNN}, a Quantum Tensor Network-Based
Recurrent Quantum Neural Network that integrates tensor-network-inspired
nearest-neighbor interactions with recurrent quantum processing. The
model sequentially encodes input features while propagating quantum
memory across recurrent steps and incorporates QSMOTE for class
balancing. QTN-RQNN is evaluated on four heterogeneous fault-diagnosis
datasets SPID, CWRUBD, EFDD, and SFDD and compared with QNN,
QTN-QNN, and RQNN baselines through classification, ablation,
multi-seed, convergence, and quantum-resource analyses. Results show
dataset-dependent predictive behavior, indicating that increased
quantum architectural complexity does not universally improve
classification. On SPID, QTN-RQNN improves the RQNN F1-score from
$0.3077$ to $0.4516$ and ROC-AUC from $0.4485$ to $0.6970$, while
the remaining datasets exhibit different performance-resource
trade-offs and sensitivity to recurrent timesteps and circuit depth.
Across all datasets, QTN-RQNN reduces the two-qubit-gate count from
32 to 24 per sample relative to RQNN, corresponding to a $25\%$
reduction without increasing the four-qubit requirement; the same
relative reduction is obtained for QTN-QNN over QNN. These findings
establish QTN-RQNN as a resource-conscious recurrent quantum
architecture and highlight the need to jointly consider predictive
performance and circuit-resource requirements for near-term quantum
fault diagnosis.
\end{abstract}


% \fi
\IEEEpeerreviewmaketitle

\section{Introduction}
\label{Sec1}

Predictive maintenance and intelligent condition monitoring are increasingly
important in modern industrial systems, where early fault detection can
reduce unplanned downtime and improve operational reliability and safety
\cite{surucu2023condition,vashishtha2024roadmap}. The widespread use of
sensing and Internet-of-Things technologies has enabled large-scale
collection of vibration, image, and multivariate sensor data, motivating
machine learning (ML) and deep learning (DL) methods for automated fault
diagnosis \cite{surucu2023condition,zhao2019deep}. However, industrial data
are often high-dimensional, nonlinear, noisy, and imbalanced, while complex
learning models can impose substantial computational requirements
\cite{vashishtha2024roadmap}.

Quantum machine learning (QML) offers an alternative computational framework
based on quantum feature representations and parameterized quantum circuits
\cite{havlivcek2019supervised,schuld2021machine,cerezo2021variational}.
Recent studies have explored QML for industrial classification and predictive
maintenance \cite{ajagekar2020quantum,ajagekar2021quantum,
diedrich2024industrial}. Nevertheless, limited qubit availability, circuit
depth, hardware noise, and optimization difficulties remain major constraints
on noisy intermediate-scale quantum (NISQ) devices
\cite{cerezo2021variational,preskill2018quantum}. Increasing variational
circuit size can also enlarge the optimization space and exacerbate
trainability difficulties \cite{cerezo2021variational,mcclean2018barren}.
Resource-conscious quantum architectures are therefore required to balance
representational capability against circuit complexity.

Quantum tensor networks (QTNs) provide one approach to structured quantum
learning. Matrix product states (MPS), tree tensor networks (TTNs), and
multiscale entanglement renormalization ansatz (MERA) have been investigated
as structured quantum models for supervised learning
\cite{huggins2019towards,bhatia2019matrix,huang2021variational,
wall2021tree,lazzarin2022multiclass,rieser2023tensor}. Hybrid
tensor-network-variational approaches further demonstrate the potential of
structured connectivity for QML \cite{konar2025tensor}. These architectures,
however, are predominantly feedforward and do not generally maintain
persistent quantum memory across sequential feature segments.

Recurrent Quantum Neural Networks (RQNNs), in contrast, process information
sequentially while propagating quantum information and reusing parameterized
operations across recurrent steps
\cite{bausch2020recurrent,takaki2021learning,li2023quantum,
huang2026fully}. Such recurrence can limit parameter growth with sequence
length and provide memory of preceding inputs, but existing RQNNs generally
do not explicitly impose a tensor-network-inspired topology on the recurrent
quantum transformation. A mathematical relationship between tensor networks
and recurrent representations has been demonstrated by Wu \textit{et al.}
\cite{wu2023tensor}; however, that formulation concerns tensorial recurrent
representations of quantum many-body states rather than a recurrent
parameterized quantum-circuit classifier. Thus, structured QTN correlation
learning and recurrent quantum information processing remain largely
separated at the architectural level.

\begin{figure*}[!t]
    \centering
    \includegraphics[width=\textwidth]{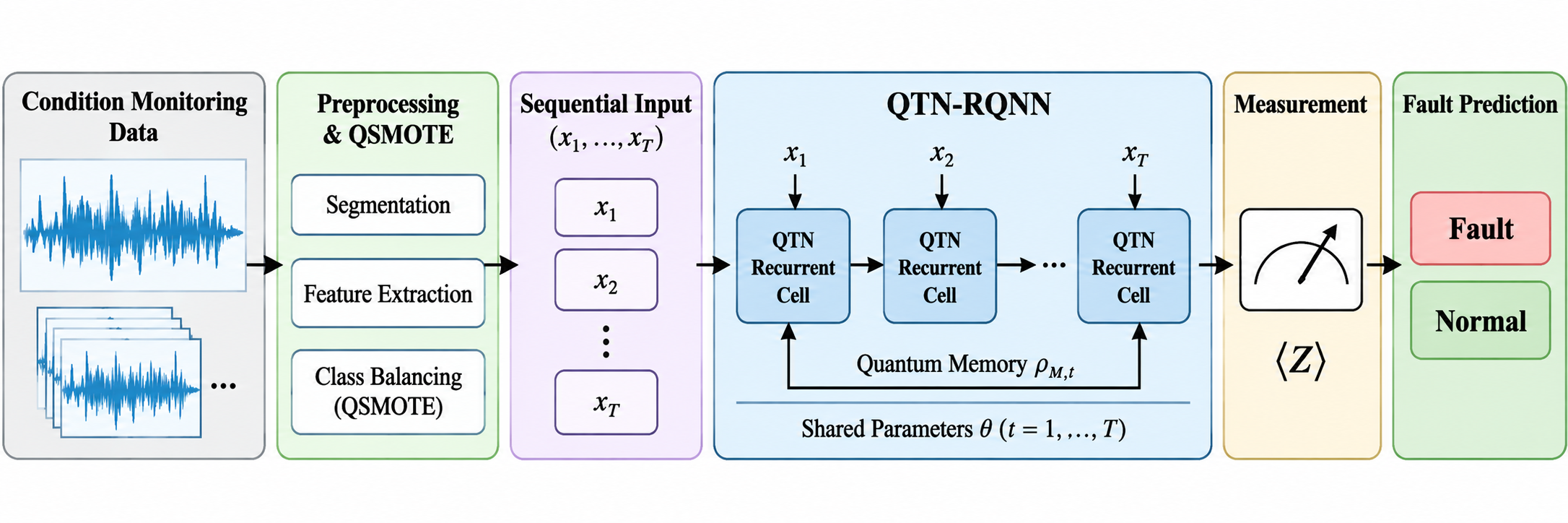}
    \caption{Compact overview of the proposed QTN-RQNN framework.
    Preprocessed and QSMOTE-balanced features are processed sequentially
    by shared QTN-based recurrent cells while quantum memory is propagated
    across recurrent steps. Final measurements are used for fault
    classification.}
    \label{fig:qtn_rqnn_framework}
\end{figure*}

To bridge this gap, we propose \textbf{QTN-RQNN}, a Quantum Tensor
Network-Based Recurrent Quantum Neural Network that incorporates
tensor-network-inspired local interactions directly within a recurrent
quantum cell. As illustrated in Fig.~\ref{fig:qtn_rqnn_framework},
preprocessed and class-balanced features are partitioned into sequential
inputs and processed using shared QTN-RQNN parameters, while quantum memory
is propagated across recurrent steps. This construction combines structured
intra-cell interactions with inter-step information propagation and parameter
reuse, rather than cascading an independent QTN feature extractor with an
RQNN.

\subsection{Main Contributions}

The main contributions of this work are summarized as follows:
\begin{itemize}

\item We propose \textbf{QTN-RQNN}, which integrates
tensor-network-inspired structured interactions directly into a recurrent
quantum architecture with propagated quantum memory and shared parameters.

\item We develop a resource-conscious recurrent quantum cell that combines
structured local interactions with recurrent parameter reuse, enabling
sequential feature processing without increasing the qubit requirement with
the number of feature segments.

\item We isolate the effects of tensor-network structuring and recurrence
through comparisons with QNN, QTN-QNN, and RQNN baselines and through
timestep, circuit-depth, class-balancing, and robustness analyses.

\item We evaluate the framework on four heterogeneous fault-diagnosis
datasets and jointly assess predictive performance and quantum-resource
requirements, including qubit, parameter, and entangling-gate costs.

\end{itemize}

\section{Background and Related Work}
\label{sec:related_work}

\subsection{Quantum Tensor Networks for Machine Learning}

Tensor networks (TNs) represent high-dimensional quantum states through
structured networks of lower-order tensors, with prominent forms including
matrix product states (MPS), tree tensor networks (TTNs), projected
entangled-pair states (PEPS), and the multiscale entanglement
renormalization ansatz (MERA). Their structured correlation patterns and
connections to quantum circuits have motivated their use in quantum machine
learning (QML) \cite{huggins2019towards,rieser2023tensor}. Bhatia
\textit{et al.} \cite{bhatia2019matrix} demonstrated MPS-based quantum
classification, while subsequent studies investigated variational QTN,
TTN, and TTN/MERA classifiers
\cite{huang2021variational,wall2021tree,lazzarin2022multiclass}.
More recently, TR-QNet combined tensor-ring representations with
variational quantum learning \cite{konar2025tensor}. Collectively, these
studies demonstrate the potential of tensor-network-inspired structures
for controlling circuit connectivity and representational complexity;
however, the resulting learning architectures are predominantly
feedforward and do not maintain persistent quantum memory across
sequential inputs \cite{rieser2023tensor}.

\subsection{Recurrent Quantum Learning and TN-Recurrent Connections}

Recurrent quantum neural networks (RQNNs) provide a complementary approach
by processing sequential information while reusing parameterized quantum
operations across timesteps. Bausch \cite{bausch2020recurrent} introduced
an early RQNN for sequence learning, followed by variational recurrent
quantum formulations for temporal data \cite{takaki2021learning}.
Li \textit{et al.} \cite{li2023quantum} subsequently developed a
hardware-efficient QRNN based on staggered recurrent blocks, while
Huang and Yi \cite{huang2026fully} investigated a fully quantum-native
recurrent architecture for sequential learning on NISQ hardware. These
models establish recurrent quantum processing as a mechanism for
parameter reuse and information propagation, but their recurrent cells
do not explicitly impose tensor-network-inspired connectivity. A theoretical connection between tensor networks and recurrent
representations was established by Wu \textit{et al.}
\cite{wu2023tensor}, who showed that MPS representations can be mapped to
recurrent neural-network forms and generalized to tensorial recurrent
networks. However, these are classical recurrent representations of
quantum many-body states rather than recurrent parameterized
quantum-circuit classifiers. From another perspective, Shin
\textit{et al.} \cite{shin2024dequantizing} analyzed variational QML
through tensor-network representations, emphasizing that tensor-network
structure alone does not establish computational advantage and motivating
architectures whose structural constraints provide measurable
performance or resource benefits.

\subsection{Positioning of QTN-RQNN}

Existing QTN-based classifiers primarily exploit structured connectivity
within feedforward quantum models, whereas RQNNs exploit recurrent memory
and parameter reuse without explicitly imposing a tensor-network-inspired
topology on the recurrent transformation. Although tensor-network and
recurrent representations are theoretically related
\cite{wu2023tensor}, their integration within a shared recurrent
parameterized quantum cell for supervised learning remains comparatively
unexplored. QTN-RQNN addresses this architectural gap by incorporating
tensor-network-inspired local interactions directly into the recurrent
quantum transformation. The same structured cell is reused across
sequential feature segments while quantum memory propagates information
between recurrent steps. Thus, the architecture combines structured
intra-cell connectivity with inter-step recurrent parameter sharing.
Table~\ref{tab:related_work} summarizes its positioning relative to
representative QTN and recurrent quantum-learning approaches.

\begin{table*}[!t]
\centering
\caption{Comparison of QTN-RQNN with representative tensor-network and
recurrent quantum-learning architectures.}
\label{tab:related_work}
\resizebox{\textwidth}{!}{%
\begin{tabular}{lccccc}
\hline
\textbf{Method} &
\textbf{Quantum} &
\textbf{TN Structure} &
\textbf{Recurrent} &
\textbf{Quantum Memory} &
\textbf{Parameter Sharing} \\
\hline
MPS Classifier \cite{bhatia2019matrix}
& \checkmark & \checkmark & -- & -- & -- \\

VQTN \cite{huang2021variational}
& \checkmark & \checkmark & -- & -- & -- \\

TTN Classifier \cite{wall2021tree}
& \checkmark & \checkmark & -- & -- & -- \\

TTN/MERA \cite{lazzarin2022multiclass}
& \checkmark & \checkmark & -- & -- & -- \\

RQNN \cite{bausch2020recurrent}
& \checkmark & -- & \checkmark & \checkmark & \checkmark \\

VQRNN \cite{takaki2021learning}
& \checkmark & -- & \checkmark & \checkmark & \checkmark \\

Hardware-Efficient QRNN \cite{li2023quantum}
& \checkmark & -- & \checkmark & \checkmark & \checkmark \\

Tensorial RNN \cite{wu2023tensor}
& -- & \checkmark & \checkmark & -- & \checkmark \\

Quantum-Native RNN \cite{huang2026fully}
& \checkmark & -- & \checkmark & \checkmark & \checkmark \\

\textbf{QTN-RQNN (Proposed)}
& \checkmark & \checkmark & \checkmark & \checkmark & \checkmark \\
\hline
\end{tabular}%
}
\end{table*}

% ============================================================
\section{Proposed QTN-RQNN Methodology}
\label{Sec3}
% ============================================================

This section presents the proposed Quantum Tensor Network-Based
Recurrent Quantum Neural Network (QTN-RQNN). The framework is designed
for sequential fault classification using a compact quantum register
in which a subset of qubits encodes the current input segment while
the remaining qubits retain recurrent quantum memory. The central
component is a tensor-network-inspired recurrent cell constructed from
local trainable rotations and nearest-neighbor parameterized
interactions. The same cell parameters are reused across recurrent
steps, allowing sequential information processing without increasing
the number of trainable quantum parameters with sequence length.

\subsection{Problem Formulation and Overall Framework}
\label{subsec:problem}

Consider a binary fault-classification dataset
\begin{equation}
\mathcal{D}
=
\left\{
(\mathbf{x}^{(i)},y^{(i)})
\right\}_{i=1}^{N},
\qquad
y^{(i)}\in\{0,1\},
\label{eq:dataset}
\end{equation}
where $\mathbf{x}^{(i)}$ denotes the feature representation of the
$i$th observation and $y^{(i)}$ indicates the normal or fault class.
After dataset-specific preprocessing, each observation is represented
by an eight-dimensional vector. Class balancing is performed on the
training partition using QSMOTE, while validation and test samples
remain untouched. Complete preprocessing and QSMOTE procedures are
provided in the Supplementary Material. For recurrent processing, the eight-dimensional representation is
organized into a sequence
\begin{equation}
\mathbf{x}
=
(\mathbf{x}_{1},\mathbf{x}_{2},\ldots,\mathbf{x}_{T}),
\label{eq:sequence_general}
\end{equation}
where $\mathbf{x}_{t}$ denotes the feature segment presented at
recurrent step $t$. In the default configuration used in the main
experiments, $T=4$, yielding
\begin{equation}
\mathbf{x}_{t}
=
(x_{t,1},x_{t,2}),
\qquad t=1,\ldots,4.
\label{eq:sequence_default}
\end{equation}

QTN-RQNN employs four qubits divided into an input register
$I=\{q_0,q_1\}$ and a memory register $M=\{q_2,q_3\}$. At every
recurrent step, the current two-dimensional feature segment is encoded
on the input register and jointly processed with the memory state
propagated from the preceding step. After the QTN transformation, the
input subsystem is discarded and the reduced state of the memory
subsystem is retained for the next step. The same trainable QTN
parameters are shared for all $T$ recurrent steps.

\subsection{Sequential Quantum Feature Encoding}
\label{subsec:encoding}

Before quantum encoding, the processed features are scaled to
$[0,1]$. At recurrent step $t$, the current segment
$\mathbf{x}_{t}=(x_{t,1},x_{t,2})$ is encoded on $q_0$ and $q_1$
through angle encoding. Starting from $|00\rangle_I$, the input state
is
\begin{equation}
|\phi_t\rangle_I
=
\left[
R_Y(\pi x_{t,1})
\otimes
R_Y(\pi x_{t,2})
\right]
|00\rangle_I,
\label{eq:encoding}
\end{equation}
where
\begin{equation}
R_Y(\alpha)
=
\exp\left(-i\frac{\alpha}{2}Y\right).
\end{equation}
The factor $\pi$ maps the normalized feature range to rotation angles
over $[0,\pi]$. The memory register is initialized before processing the first
segment as
\begin{equation}
\rho_{M,0}
=
|00\rangle\langle00|.
\label{eq:memory_initial}
\end{equation}
For $t>1$, the memory register instead contains the reduced state
obtained from the preceding recurrent step. The joint input-memory
state entering the recurrent cell is therefore
\begin{equation}
\rho_{IM,t}
=
|\phi_t\rangle\langle\phi_t|_I
\otimes
\rho_{M,t-1}.
\label{eq:joint_input}
\end{equation}
This construction allows every feature segment to interact with a
quantum state carrying information from previously processed
segments.

\subsection{Quantum Tensor Network-Based Recurrent Cell}
\label{subsec:qtn_cell}

\begin{figure*}[!t]
    \centering
    \includegraphics[width=\textwidth]
    {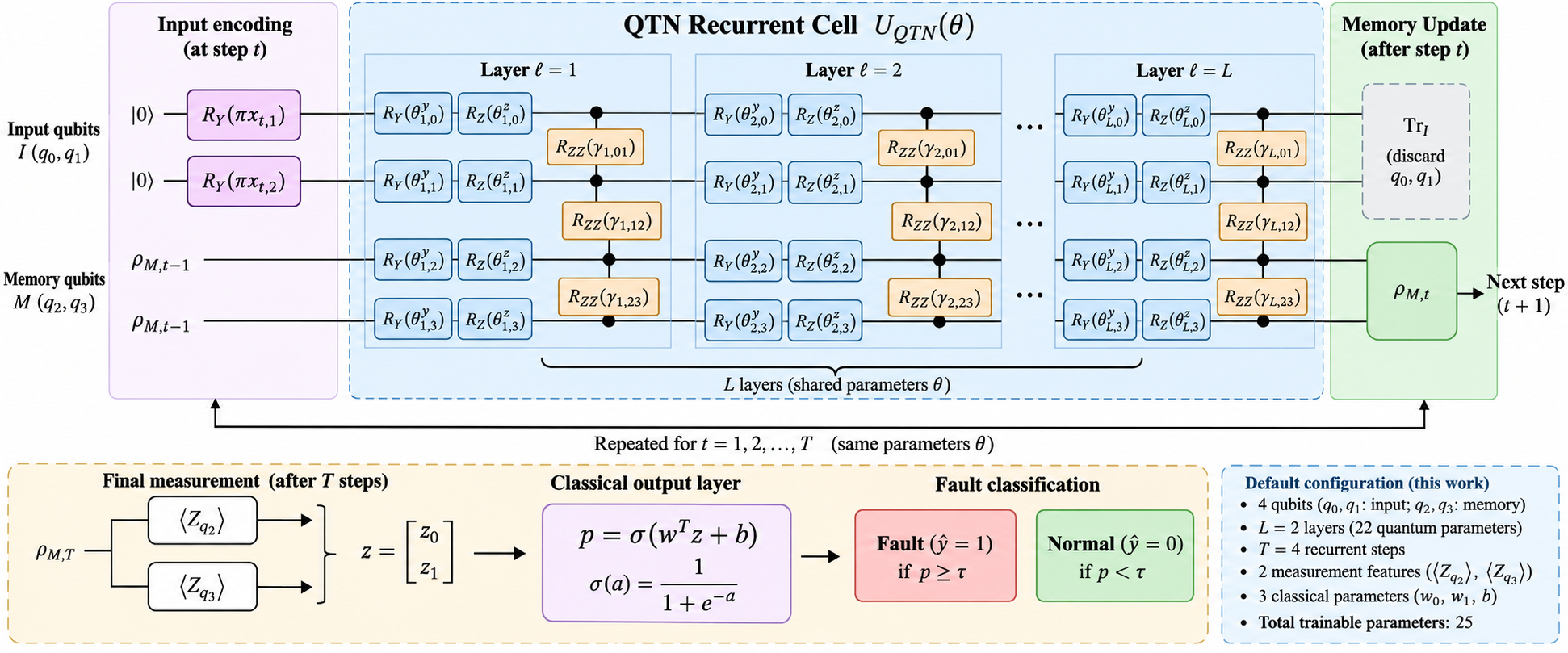}
    \caption{Circuit-level structure of the proposed QTN-RQNN.
    Qubits $q_0$ and $q_1$ encode the current feature segment through
    $R_Y(\pi x_{t,1})$ and $R_Y(\pi x_{t,2})$, while $q_2$ and $q_3$
    form the recurrent quantum-memory register. Each QTN layer applies
    trainable $R_Y$ and $R_Z$ rotations followed by parameterized
    nearest-neighbor $R_{ZZ}$ interactions on $(q_0,q_1)$,
    $(q_2,q_3)$, and $(q_1,q_2)$. After $L$ layers, the input
    subsystem is traced out and the resulting memory state
    $\rho_{M,t}$ is propagated to the next recurrent step. The same
    QTN parameters are reused for $t=1,\ldots,T$. After the final
    step, local $Z$ measurements of the memory qubits are supplied to
    a classical sigmoid output layer.}
    \label{fig:qtn_circuit}
\end{figure*}

The central architectural component is the QTN recurrent cell shown
at circuit level in Fig.~\ref{fig:qtn_circuit}. Rather than using
dense or ring-like entanglement, the cell employs a nearest-neighbor
brickwork structure inspired by one-dimensional tensor-network
connectivity. This restricts the two-qubit interactions while
preserving coupling between the input and memory subsystems. For a four-qubit register, each QTN layer $\ell$ first applies
independent trainable $R_Y$ and $R_Z$ rotations to every qubit:
\begin{equation}
U_{\mathrm{loc}}^{(\ell)}
=
\prod_{q=0}^{3}
R_Z^{(q)}(\theta_{\ell,q}^{z})
R_Y^{(q)}(\theta_{\ell,q}^{y}).
\label{eq:qtn_local}
\end{equation}
The local rotations provide two trainable parameters per qubit and
allow the state of each qubit to adapt before the structured
two-qubit interactions. The entangling stage follows the nearest-neighbor bonds
\begin{equation}
\mathcal{E}
=
\{(q_0,q_1),(q_1,q_2),(q_2,q_3)\}.
\label{eq:qtn_edges}
\end{equation}
Consistent with the implemented brickwork circuit, the interactions
are separated into an even-bond stage
\begin{equation}
U_{\mathrm{even}}^{(\ell)}
=
R_{ZZ}^{(0,1)}(\gamma_{\ell,01})
R_{ZZ}^{(2,3)}(\gamma_{\ell,23}),
\label{eq:even}
\end{equation}
followed by the odd-bond stage
\begin{equation}
U_{\mathrm{odd}}^{(\ell)}
=
R_{ZZ}^{(1,2)}(\gamma_{\ell,12}),
\label{eq:odd}
\end{equation}
where
\begin{equation}
R_{ZZ}(\gamma)
=
\exp\left(
-i\frac{\gamma}{2}Z\otimes Z
\right).
\end{equation}

The complete $\ell$th QTN layer is therefore
\begin{equation}
U_{\mathrm{QTN}}^{(\ell)}
=
U_{\mathrm{odd}}^{(\ell)}
U_{\mathrm{even}}^{(\ell)}
U_{\mathrm{loc}}^{(\ell)}.
\label{eq:qtn_layer}
\end{equation}
For circuit depth $L$, the recurrent-cell unitary becomes
\begin{equation}
U_{\mathrm{QTN}}(\boldsymbol{\theta})
=
U_{\mathrm{QTN}}^{(L)}
\cdots
U_{\mathrm{QTN}}^{(2)}
U_{\mathrm{QTN}}^{(1)}.
\label{eq:qtn_full}
\end{equation}

Each layer contains eight local rotation parameters and three
$R_{ZZ}$ interaction parameters. Hence, for $q=4$,
\begin{equation}
N_{\theta}
=
L[2q+(q-1)]
=
11L.
\label{eq:qtn_parameter_count}
\end{equation}
The default configuration uses $L=2$, giving $22$ trainable quantum
parameters.
Importantly, the bond $(q_1,q_2)$ couples the input and memory
subsystems. Consequently, information contained in the current
feature segment can modify the memory state before the input
subsystem is discarded. The nearest-neighbor construction also avoids
the additional closing interaction required by the ring-entangled
RQNN baseline.

\subsection{Recurrent Quantum Memory Propagation}
\label{subsec:memory}

At step $t$, the joint input-memory state in
\eqref{eq:joint_input} is transformed according to
\begin{equation}
\widetilde{\rho}_{IM,t}
=
U_{\mathrm{QTN}}(\boldsymbol{\theta})
\rho_{IM,t}
U_{\mathrm{QTN}}^{\dagger}(\boldsymbol{\theta}).
\label{eq:qtn_evolution}
\end{equation}
The input subsystem is then removed by partial trace,
\begin{equation}
\rho_{M,t}
=
\operatorname{Tr}_{I}
\left[
\widetilde{\rho}_{IM,t}
\right],
\label{eq:memory_update}
\end{equation}
and the resulting two-qubit density operator is propagated to the
next recurrent step. Combining the encoding, QTN transformation, and partial trace defines
the recurrent map
\begin{equation}
\rho_{M,t}
=
\mathcal{F}_{\boldsymbol{\theta}}
\left(
\mathbf{x}_{t},
\rho_{M,t-1}
\right),
\label{eq:recurrent_map}
\end{equation}
where the same $\boldsymbol{\theta}$ is used for every $t$. Parameter
sharing is therefore temporal: increasing $T$ increases the number of
cell applications but does not introduce a new QTN parameter set for
each step. The recurrence is implemented as a sequence of
input-conditioned quantum channels acting on the memory subsystem.

\subsection{Measurement and Fault Classification}
\label{subsec:measurement}

After processing all $T$ segments, classification is performed using
the final memory state $\rho_{M,T}$. The implemented quantum readout
consists of the local Pauli-$Z$ expectation values of the two memory
qubits,
\begin{equation}
z_0
=
\operatorname{Tr}
\left[
\rho_{M,T}
(Z\otimes I)
\right],
\qquad
z_1
=
\operatorname{Tr}
\left[
\rho_{M,T}
(I\otimes Z)
\right].
\label{eq:measurement}
\end{equation}
Defining
\begin{equation}
\mathbf{z}=[z_0,z_1]^{\top},
\end{equation}
the two quantum observables are mapped to a fault probability through
a classical logistic output layer,
\begin{equation}
\widehat{p}
=
\sigma
\left(
\mathbf{w}^{\top}\mathbf{z}+b
\right),
\qquad
\sigma(a)=\frac{1}{1+\exp(-a)}.
\label{eq:probability}
\end{equation}
Here, $\mathbf{w}\in\mathbb{R}^{2}$ and $b\in\mathbb{R}$ introduce
three classical trainable parameters. Thus, for the default $L=2$
configuration, QTN-RQNN contains
\begin{equation}
N_{\mathrm{train}}
=
22+2+1
=
25
\label{eq:total_params}
\end{equation}
trainable parameters in total.

\subsection{Training Objective and Optimization}
\label{subsec:training}

The quantum parameters $\boldsymbol{\theta}$ and classical output
parameters $(\mathbf{w},b)$ are optimized jointly. To account for
class imbalance remaining after preprocessing, training minimizes a
class-weighted binary cross-entropy objective,
\begin{equation}
\mathcal{L}
=
-\frac{1}{N_{\mathrm{tr}}}
\sum_{i=1}^{N_{\mathrm{tr}}}
\left[
w_1y_i\log\widehat{p}_i
+
w_0(1-y_i)\log(1-\widehat{p}_i)
\right],
\label{eq:weighted_bce}
\end{equation}
where $w_0$ and $w_1$ denote class weights computed from the
optimization subset. The implementation uses COBYLA, a derivative-free optimizer, with
multiple parameter initializations. Model selection is performed using
validation F1-score. A classification threshold $\tau$ is subsequently
selected on the validation set and held fixed for test evaluation:
\begin{equation}
\widehat{y}
=
\mathbb{I}
\left[
\widehat{p}\geq\tau
\right].
\label{eq:decision}
\end{equation}
This separates parameter optimization from final test evaluation and
allows the operating threshold to account for the imbalanced
fault-detection setting. Further optimization and threshold-selection
details are given in the Supplementary Material.

\subsection{Quantum-Resource Characteristics}
\label{subsec:complexity}

The structured connectivity is intended to reduce two-qubit
interactions without changing the qubit requirement. For $q$ qubits
and QTN depth $L$, each cell application contains
\begin{equation}
N_{\mathrm{2Q}}^{\mathrm{cell}}
=
L(q-1)
\label{eq:qtn_2q_cell}
\end{equation}
parameterized two-qubit gates. Over $T$ recurrent steps,
\begin{equation}
N_{\mathrm{2Q}}^{\mathrm{QTN-RQNN}}
=
TL(q-1).
\label{eq:qtn_2q_total}
\end{equation}
For the default $q=4$, $L=2$, and $T=4$ configuration, this gives
24 two-qubit gates per sample. The conventional recurrent baseline uses a four-edge ring of
two-qubit interactions in every layer, yielding
\begin{equation}
N_{\mathrm{2Q}}^{\mathrm{RQNN}}
=
TLq
=
32.
\label{eq:rqnn_2q_total}
\end{equation}
Thus, at the logical-circuit level, the QTN connectivity reduces the
two-qubit interaction count by $25\%$ while retaining the same
four-qubit register and recurrent processing length. These counts are
pre-transpilation logical resources; complete parameter and gate-count
derivations are provided in the Supplementary Material.

\section{Experimental Results}\label{SecIV}

%\color{blue}
\subsection{Benchmark Datasets}
\label{sec:datasets}

The proposed QTN-RQNN is evaluated on four publicly available benchmarks spanning
photovoltaic inspection, bearing monitoring, engine fault detection, and sensor
fault detection, as summarized in Table~\ref{tab:datasets}. SPID
\cite{afroz2023solar} contains RGB images of six solar-panel conditions, whereas
CWRUBD \cite{cwru_kaggle} provides vibration-based bearing measurements and EFDD
\cite{engine_failure_kaggle} contains multivariate engine-condition data. SFDD
\cite{sfdd_kaggle} consists of timestamped sensor measurements; in our
implementation, eight consecutive readings form one sequential sample, with
normal/fault labels determined from abnormal local deviations and sampling-gap
events. For consistency, all experiments are formulated as binary
classification, with labels mapped to $\{0,1\}$. QSMOTE is applied only to the
training partition, while the original test distribution is retained.

\begin{table}[t]
\centering
\caption{Datasets used for QTN-RQNN evaluation.}
\label{tab:datasets}
\small
%\resizebox{\columnwidth}{!}{
\begin{tabular}{lccc}
\toprule
\textbf{Dataset} & \textbf{Domain} & \textbf{Task} & \textbf{Data Type}\\
\midrule
SPID \cite{afroz2023solar} & Solar panel & Binary & Images\\
CWRUBD \cite{cwru_kaggle} & Bearing & Binary & Vibration\\
EFDD \cite{engine_failure_kaggle} & Engine & Binary & Sensor\\
SFDD \cite{sfdd_kaggle} & Sensor fault & Binary & Time-series\\
\bottomrule
\end{tabular}%}
\end{table}

\subsection{Data Preparation}
\label{sec:preprocessing}

Dataset-specific transformations are performed before quantum encoding. For
SPID, ResNet50 is used to obtain image embeddings, which are subsequently
compressed through PCA. CWRUBD vibration signals are represented by nine
time-domain statistics, including RMS, skewness, kurtosis, crest factor, and
form factor. For EFDD, categorical variables are numerically encoded and
non-predictive attributes, such as timestamps, are removed. The resulting
representations of SPID, CWRUBD, and EFDD are reduced to eight components using
PCA fitted on the training data. SFDD is processed differently because its temporal structure is directly
exploited by the recurrent model. After chronological organization of the
sensor measurements, each instance is formed from eight consecutive readings;
hence, no PCA is required for this dataset. For all benchmarks, the final
features are scaled using parameters estimated from the training partition
only. The resulting eight-dimensional inputs are subsequently arranged as four
two-feature segments for the default QTN-RQNN configuration, enabling sequential
encoding with two input qubits.

\subsection{Comparative Models}

To isolate the contributions of tensor-network structuring and recurrent
processing, QTN-RQNN is compared with three architectural variants: QNN,
QTN-QNN, and RQNN. QNN provides a non-recurrent variational baseline,
QTN-QNN introduces the QTN-structured circuit without recurrence, and RQNN
incorporates recurrent quantum memory without QTN structuring. QTN-RQNN
combines both mechanisms. All models use the same processed inputs, data
partitions, and evaluation procedure, allowing their architectural effects on
fault classification to be examined consistently.

\subsection{Hyperparameters and Evaluation Metrics}
\label{Sec4.1}

All models are evaluated using common data partitions and preprocessing settings,
with QSMOTE restricted to the training set. Unless otherwise stated, experiments
use \texttt{random\_state=42}. The default QTN-RQNN employs four qubits (two
input and two memory qubits), QTN depth $L=2$, and processes each
eight-dimensional input over $T=4$ timesteps with two features per timestep.
Model parameters are optimized using COBYLA with a maximum of 60 iterations and
three independent restarts. Performance is quantified using accuracy, precision, recall, F1-score,
ROC-AUC, and PR-AUC, with particular emphasis on recall, F1-score, and PR-AUC
for minority-fault detection. Computational efficiency is additionally examined
through training time, number of trainable parameters, qubit count, and
single- and two-qubit gate counts. Robustness and architectural sensitivity are
assessed through optimizer-seed experiments and ablations of QSMOTE, recurrent
timesteps $T$, and QTN depth $L$.
\color{black}

\subsection{Results and Analysis}\label{Sec4.3}

The SPID experiments evaluate the proposed QTN-RQNN under an imbalanced
binary fault-classification setting. Before QSMOTE, the training partition
contained 129 samples from the majority class and 40 samples from the
minority class. QSMOTE was applied exclusively to the training partition,
resulting in a balanced training distribution of 129 samples per class,
while the test set was kept completely unchanged with 33 majority-class
and 10 minority-class samples. PCA reduced the extracted representation to
eight dimensions, retaining approximately $71.82\%$ of the cumulative
variance. Unless otherwise stated, the QTN-RQNN configuration used four
recurrent timesteps ($T=4$), two QTN layers ($L=2$), four qubits, and three
independent optimization restarts.

For CWRUBD, the training partition initially contained 184 majority-class
and 48 minority-class samples, while the untouched test set contained 46
and 12 samples, respectively. QSMOTE balanced the training partition to
184 samples per class. PCA retained eight components, accounting for
approximately $100\%$ of the variance. Unless otherwise specified,
QTN-RQNN uses $T=4$, $L=2$, four qubits, and three optimization restarts. For EFDD, the binary training partition initially contains 206 majority-class and 40 minority-class samples, while the untouched test
partition contains 52 and 10 samples, respectively. QSMOTE increases the
minority training set to 206 samples, resulting in 412 balanced training
instances. The processed features are represented using eight principal
components. Unless otherwise specified, QTN-RQNN employs $T=4$, $L=2$,
four qubits, and three optimization restarts.

The SFDD dataset provides a substantially larger window-level
time-series classification problem than the other datasets considered
in this study. Following the temporal preprocessing procedure described
previously, the training partition contains 29,325 normal and 20,772
fault windows, while the untouched test partition contains 7,332 normal
and 5,193 fault windows. To control the computational cost of the
QSMOTE stage, the majority-class training pool is capped at 400 samples
before synthetic balancing. QSMOTE subsequently produces a balanced
training representation containing 20,772 samples from each class.
Unlike the other tabular datasets, SFDD uses eight consecutive sensor
readings directly and therefore does not require PCA. Unless otherwise
specified, QTN-RQNN employs $T=4$, $L=2$, four qubits, and three
optimization restarts. 

% ============================================================
\subsection{Classification Performance}
\label{subsec:classification_performance}
% ============================================================

\begin{figure*}[!t]
    \centering

    % =========================================================
    % Row 1: ROC curves
    % =========================================================
    \begin{subfigure}[b]{0.235\textwidth}
        \centering
        \includegraphics[width=\linewidth]
        {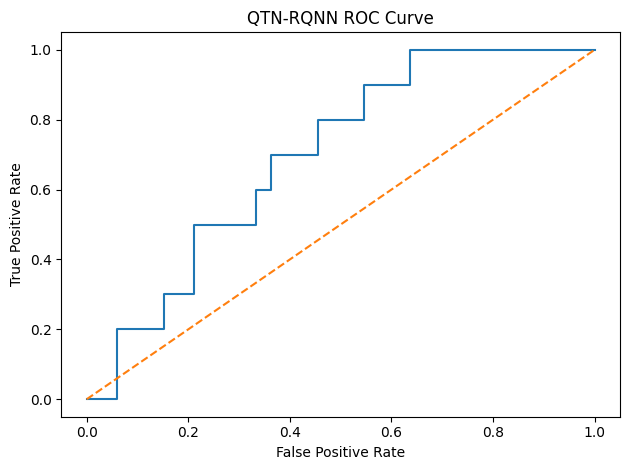}
        \caption{SPID: ROC.}
        \label{fig:spid_roc}
    \end{subfigure}
    \hfill
    \begin{subfigure}[b]{0.235\textwidth}
        \centering
        \includegraphics[width=\linewidth]
        {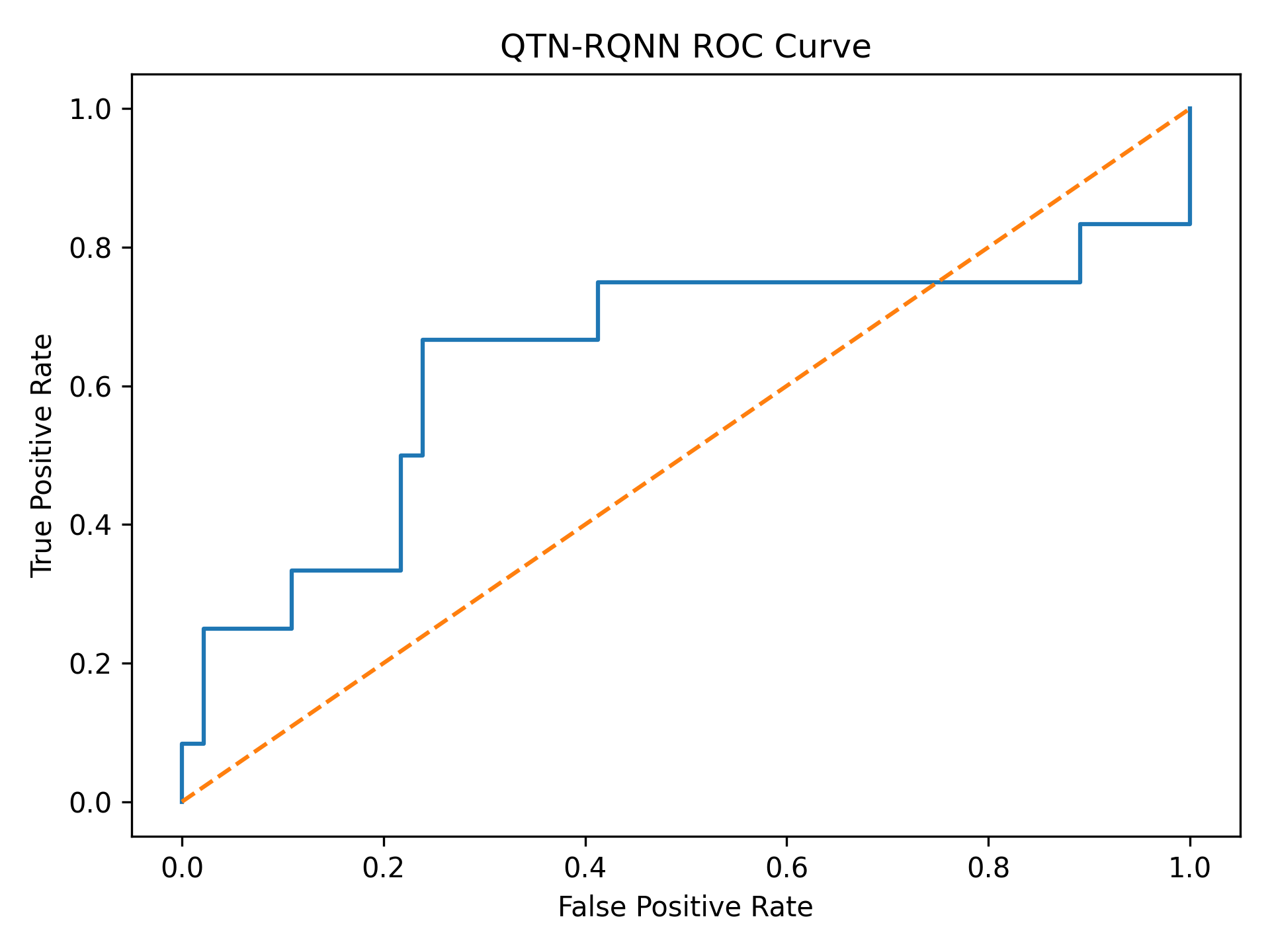}
        \caption{CWRUBD: ROC.}
        \label{fig:cwru_roc}
    \end{subfigure}
    \hfill
    \begin{subfigure}[b]{0.235\textwidth}
        \centering
        \includegraphics[width=\linewidth]
        {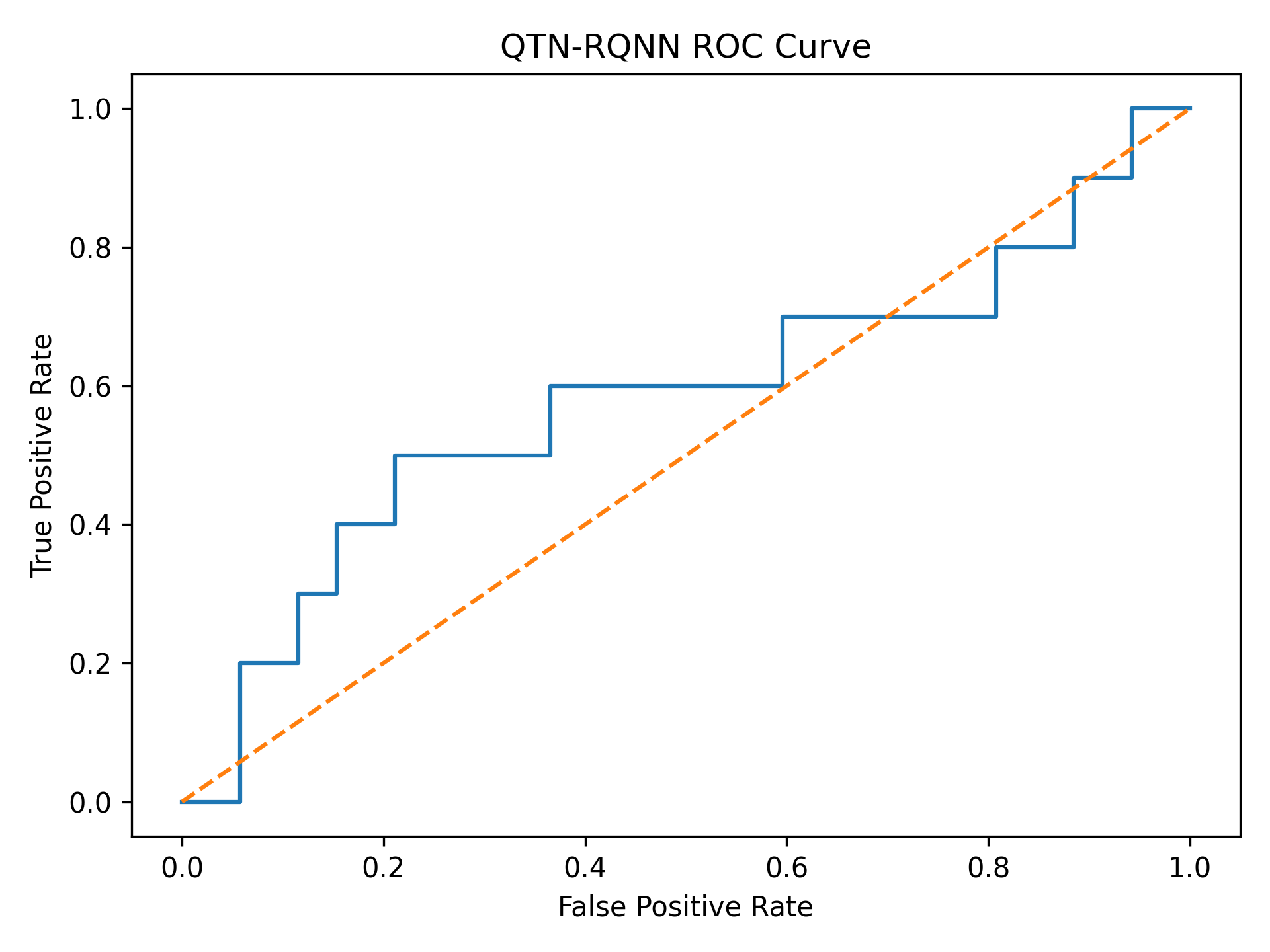}
        \caption{EFDD: ROC.}
        \label{fig:efdd_roc}
    \end{subfigure}
    \hfill
    \begin{subfigure}[b]{0.235\textwidth}
        \centering
        \includegraphics[width=\linewidth]
        {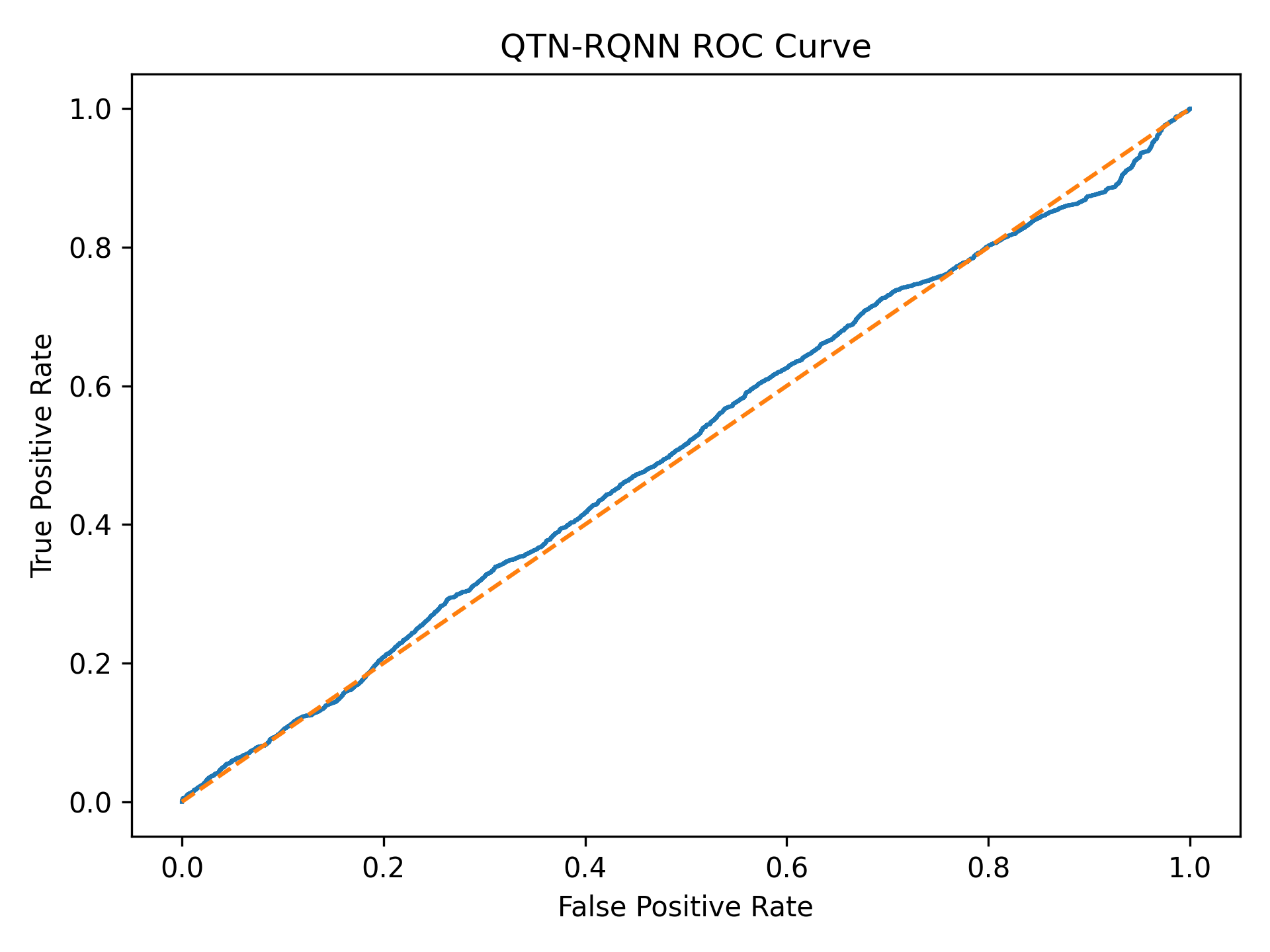}
        \caption{SFDD: ROC.}
        \label{fig:sfdd_roc}
    \end{subfigure}

    \vspace{0.5em}

    % =========================================================
    % Row 2: Precision--Recall curves
    % =========================================================
    \begin{subfigure}[b]{0.235\textwidth}
        \centering
        \includegraphics[width=\linewidth]
        {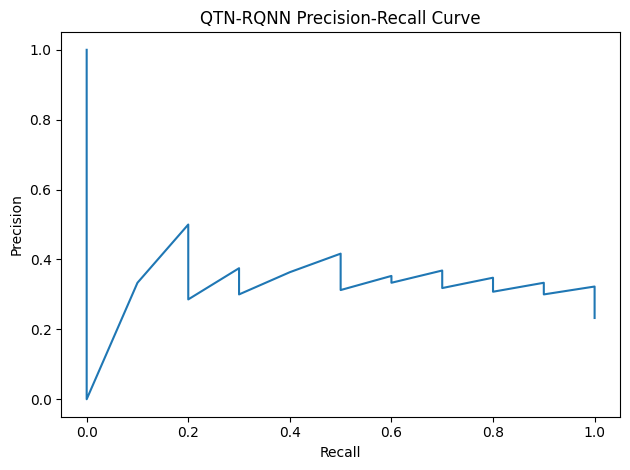}
        \caption{SPID: PR.}
        \label{fig:spid_pr}
    \end{subfigure}
    \hfill
    \begin{subfigure}[b]{0.235\textwidth}
        \centering
        \includegraphics[width=\linewidth]
        {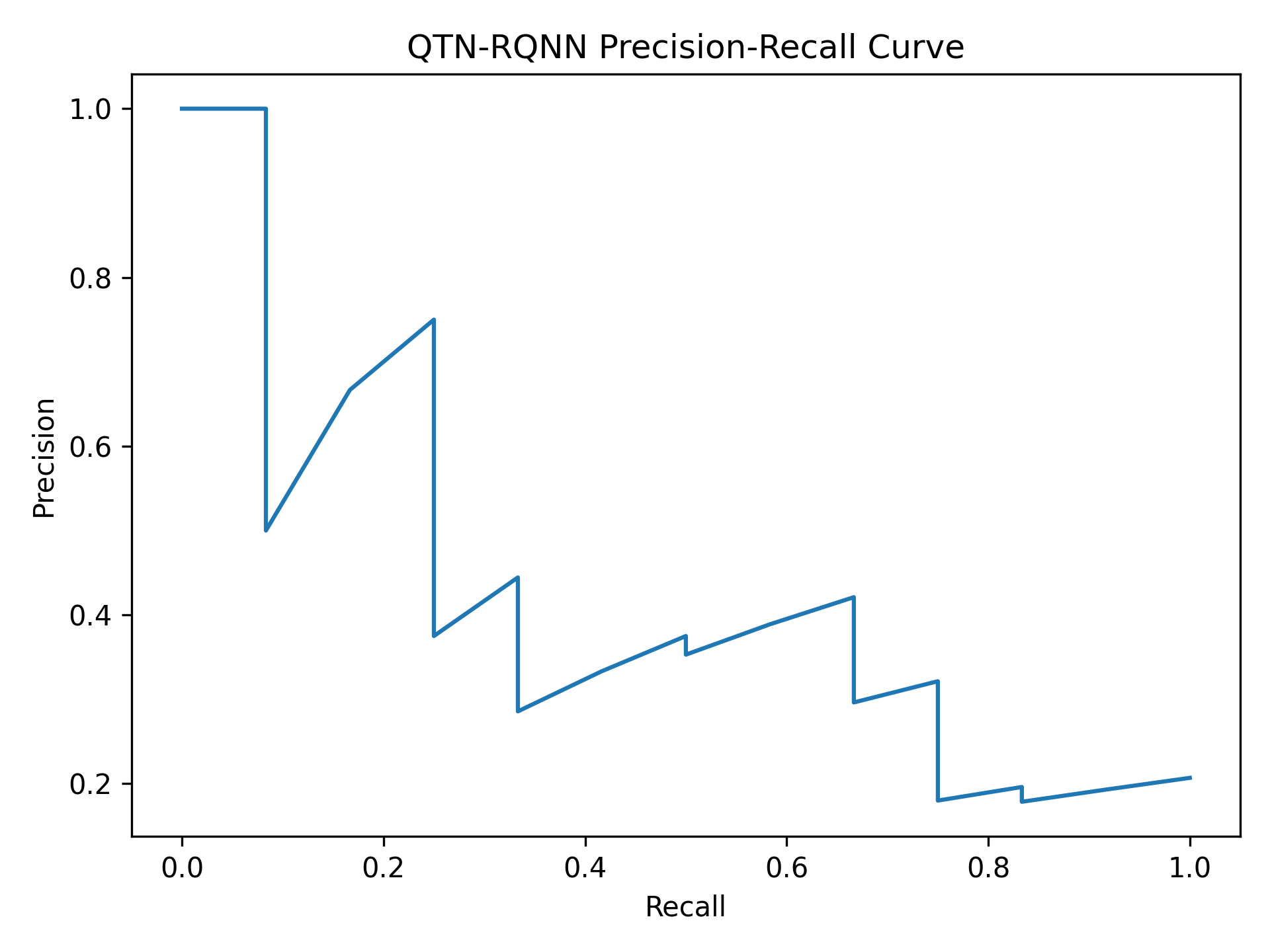}
        \caption{CWRUBD: PR.}
        \label{fig:cwru_pr}
    \end{subfigure}
    \hfill
    \begin{subfigure}[b]{0.235\textwidth}
        \centering
        \includegraphics[width=\linewidth]
        {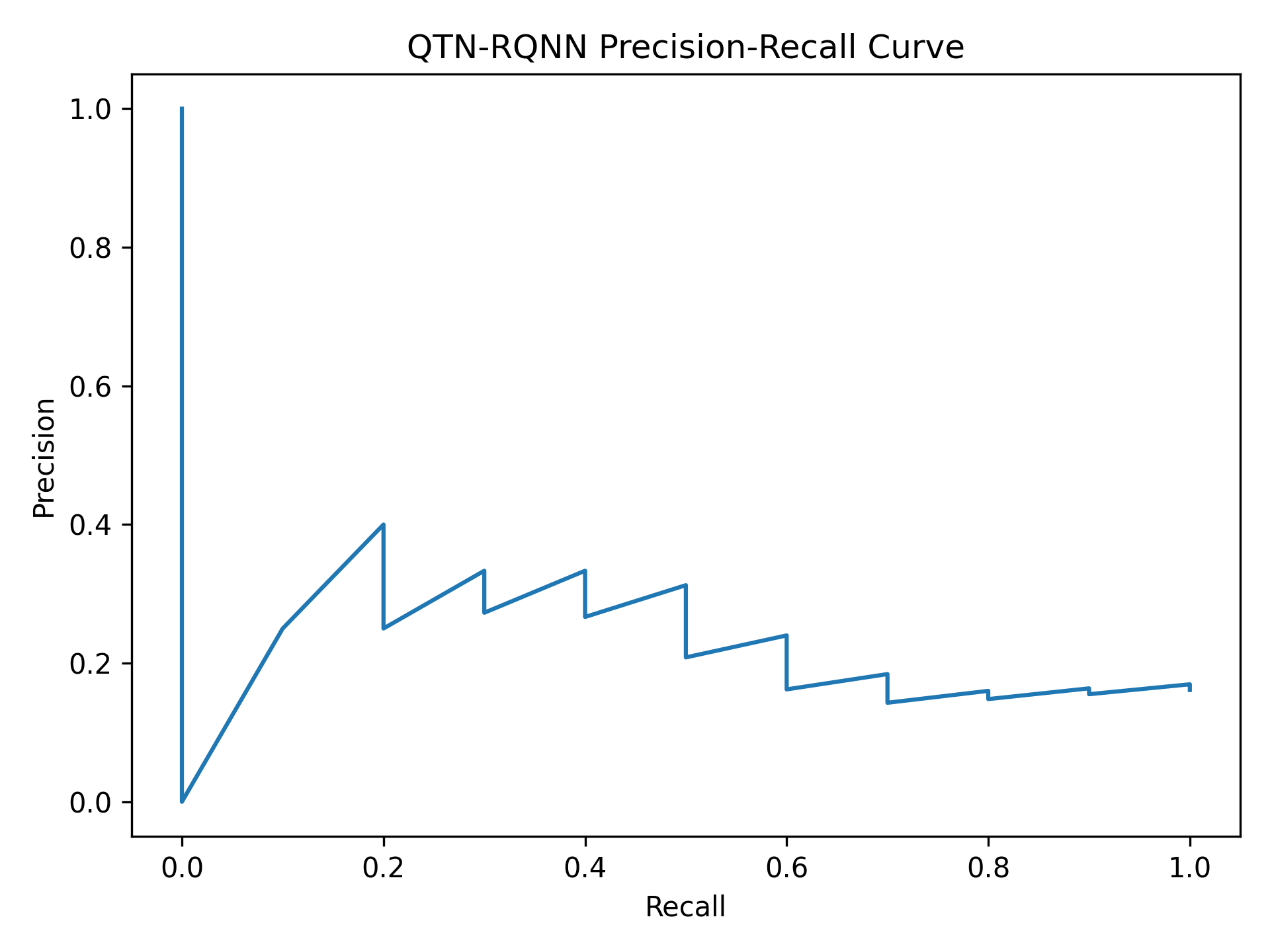}
        \caption{EFDD: PR.}
        \label{fig:efdd_pr}
    \end{subfigure}
    \hfill
    \begin{subfigure}[b]{0.235\textwidth}
        \centering
        \includegraphics[width=\linewidth]
        {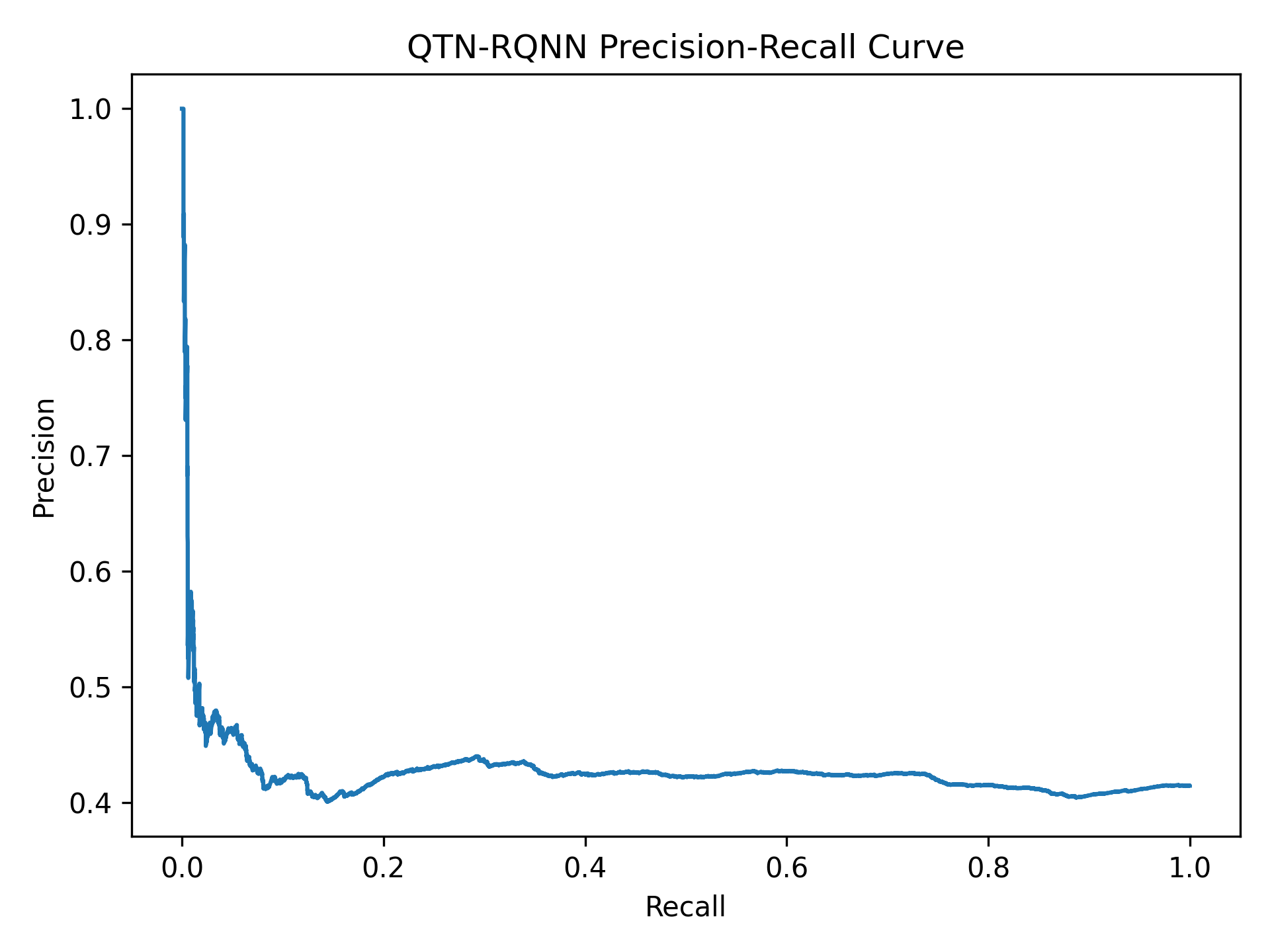}
        \caption{SFDD: PR.}
        \label{fig:sfdd_pr}
    \end{subfigure}

    \caption{Threshold-independent discrimination performance of
    QTN-RQNN across the four datasets. The first row presents the
    ROC curves for (a) SPID, (b) CWRUBD, (c) EFDD, and (d) SFDD,
    while the second row presents the corresponding precision-recall
    (PR) curves for (e) SPID, (f) CWRUBD, (g) EFDD, and (h) SFDD.}
    \label{fig:all_roc_pr}
\end{figure*}

The classification performance of QTN-RQNN across the four datasets is
summarized in Table~\ref{tab:classification_all}. The results reveal
dataset-dependent behavior, particularly in the trade-off between
minority-fault sensitivity and majority-class specificity. On SPID,
QTN-RQNN achieves an accuracy of $0.6047$, F1-score of $0.4516$,
ROC-AUC of $0.6970$, and PR-AUC of $0.3714$, while detecting $70\%$ of
the minority fault samples. On CWRUBD and EFDD, the validation-selected
operating thresholds yield perfect fault recall, although this is
accompanied by a large number of false-positive predictions. SFDD
similarly achieves a recall of $1.0000$ and an F1-score of $0.5862$,
but its ROC-AUC of $0.5093$ indicates limited score-level separation
between normal and fault windows. The threshold-independent discrimination characteristics are presented
jointly in Fig.~\ref{fig:all_roc_pr}. Among the four datasets, SPID
exhibits the strongest ROC discrimination, with a ROC-AUC of $0.6970$,
followed by CWRUBD ($0.6359$) and EFDD ($0.5808$). In contrast, the
SFDD ROC-AUC of $0.5093$ is close to random ranking. The corresponding
precision-recall curves provide a complementary assessment under class
imbalance, yielding PR-AUC values of $0.3714$, $0.4414$, $0.2547$, and
$0.4267$ for SPID, CWRUBD, EFDD, and SFDD, respectively.

\begin{table}[!t]
\centering
\caption{QTN-RQNN classification performance on the four datasets.
Thresholds ($\tau$) were selected on the validation set.}
\label{tab:classification_all}
\footnotesize
\setlength{\tabcolsep}{2.3pt}
\renewcommand{\arraystretch}{1.08}
\begin{tabular}{lccccccc}
\hline
\textbf{Dataset} &
\textbf{Acc.} &
\textbf{Prec.} &
\textbf{Rec.} &
\textbf{F1} &
\textbf{ROC} &
\textbf{PR} &
$\boldsymbol{\tau}$ \\
\hline
SPID   & 0.6047 & 0.3333 & 0.7000 & 0.4516 & 0.6970 & 0.3714 & 0.47 \\
CWRUBD & 0.2069 & 0.2069 & 1.0000 & 0.3429 & 0.6359 & 0.4414 & 0.05 \\
EFDD   & 0.1935 & 0.1667 & 1.0000 & 0.2857 & 0.5808 & 0.2547 & 0.41 \\
SFDD   & 0.4146 & 0.4146 & 1.0000 & 0.5862 & 0.5093 & 0.4267 & 0.05 \\
\hline
\end{tabular}
\end{table}

The four datasets expose two distinct aspects of QTN-RQNN
behavior. SPID demonstrates comparatively stronger score-level
discrimination and a more balanced operating point, whereas CWRUBD and
EFDD retain moderate ranking capability despite aggressive
fault-sensitive thresholds. SFDD represents the most challenging case:
although the selected threshold provides complete fault sensitivity,
the near-random ROC-AUC shows that this result is primarily
threshold-driven rather than evidence of strong intrinsic class
separation. These observations emphasize the need to consider
threshold-dependent metrics together with ROC-AUC, PR-AUC, and the
confusion matrices when evaluating QTN-RQNN under heterogeneous class
distributions.

% ============================================================
\subsection{Architecture Ablation Study}
\label{subsec:architecture_ablation}
% ============================================================

\begin{table}[!t]
\centering
\caption{Architecture ablation across the four fault-diagnosis datasets.
Time is reported in seconds.}
\label{tab:arch_ablation_all}
\footnotesize
\setlength{\tabcolsep}{1.45pt}
\renewcommand{\arraystretch}{1.03}
\begin{tabular}{llccccccc}
\hline
\textbf{Data} &
\textbf{Model} &
\textbf{Acc.} &
\textbf{Prec.} &
\textbf{Rec.} &
\textbf{F1} &
\textbf{ROC} &
\textbf{PR} &
\textbf{T(s)} \\
\hline

\multirow{4}{*}{SPID}
& QNN      & 0.6279 & 0.3750 & 0.9000 & 0.5294 & 0.7879 & 0.5812 & 9.13 \\
& QTN-QNN  & 0.6279 & 0.3125 & 0.5000 & 0.3846 & 0.6576 & 0.4584 & 13.66 \\
& RQNN     & 0.5814 & 0.2500 & 0.4000 & 0.3077 & 0.4485 & 0.2511 & 26.08 \\
& QTN-RQNN & 0.6047 & 0.3333 & 0.7000 & 0.4516 & 0.6970 & 0.3714 & 25.71 \\
\hline

\multirow{4}{*}{CWRU}
& QNN      & 0.8276 & 0.5455 & 1.0000 & 0.7059 & 0.9946 & 0.9833 & 13.13 \\
& QTN-QNN  & 0.9138 & 0.7059 & 1.0000 & 0.8276 & 0.9909 & 0.9677 & 19.30 \\
& RQNN     & 0.8448 & 0.6667 & 0.5000 & 0.5714 & 0.7029 & 0.5269 & 32.68 \\
& QTN-RQNN & 0.2069 & 0.2069 & 1.0000 & 0.3429 & 0.6359 & 0.4414 & 32.75 \\
\hline

\multirow{4}{*}{EFDD}
& QNN      & 0.2581 & 0.1250 & 0.6000 & 0.2069 & 0.3481 & 0.1343 & 12.16 \\
& QTN-QNN  & 0.2258 & 0.1607 & 0.9000 & 0.2727 & 0.5288 & 0.2052 & 18.24 \\
& RQNN     & 0.4355 & 0.1795 & 0.7000 & 0.2857 & 0.5115 & 0.2601 & 31.30 \\
& QTN-RQNN & 0.1935 & 0.1667 & 1.0000 & 0.2857 & 0.5808 & 0.2547 & 29.93 \\
\hline

\multirow{4}{*}{SFDD}
& QNN      & 0.4146 & 0.4146 & 1.0000 & 0.5862 & 0.5097 & 0.4272 & 12.25 \\
& QTN-QNN  & 0.4146 & 0.4146 & 1.0000 & 0.5862 & 0.4979 & 0.4281 & 16.08 \\
& RQNN     & 0.4146 & 0.4146 & 1.0000 & 0.5862 & 0.5093 & 0.4264 & 32.04 \\
& QTN-RQNN & 0.4146 & 0.4146 & 1.0000 & 0.5862 & 0.5093 & 0.4267 & 31.49 \\
\hline
\end{tabular}
\end{table}
To isolate the effects of tensor-network structuring and recurrence,
QNN, QTN-QNN, RQNN, and QTN-RQNN were evaluated using the same
dataset-specific preprocessing and test partitions
(Table~\ref{tab:arch_ablation_all}). The results show strongly
dataset-dependent behavior. On SPID, QNN provides the strongest
overall performance, whereas QTN-RQNN improves the recurrent RQNN in
recall ($0.4000\!\rightarrow\!0.7000$), F1-score
($0.3077\!\rightarrow\!0.4516$), and ROC-AUC
($0.4485\!\rightarrow\!0.6970$). On CWRUBD, QTN-QNN achieves the
highest accuracy and F1-score, while the default QTN-RQNN performs
substantially worse, indicating sensitivity to the recurrent
configuration. On EFDD, QTN-RQNN increases recall
($0.7000\!\rightarrow\!1.0000$) and ROC-AUC
($0.5115\!\rightarrow\!0.5808$) relative to RQNN, although RQNN
retains higher accuracy and PR-AUC. On SFDD, all architectures yield identical threshold-dependent
metrics, with ROC-AUC values close to $0.5$, indicating weak
score-level discrimination under the present configuration. The ablation does not establish a universal predictive advantage for
either recurrence or QTN structuring; their effects are
dataset-dependent. The benefit of QTN-RQNN should therefore be
considered jointly with its quantum-resource requirements, particularly
the reduction in two-qubit interactions examined subsequently.
% ============================================================
\subsection{Effect of QSMOTE}
\label{subsec:qsmote_ablation}
% ============================================================

\begin{table}[!t]
\centering
\caption{Effect of QSMOTE on QTN-RQNN across the four
fault-diagnosis datasets.}
\label{tab:qsmote_all}
\footnotesize
\setlength{\tabcolsep}{1.8pt}
\renewcommand{\arraystretch}{1.03}
\begin{tabular}{llcccccc}
\hline
\textbf{Data} &
\textbf{QSMOTE} &
\textbf{Acc.} &
\textbf{Prec.} &
\textbf{Rec.} &
\textbf{F1} &
\textbf{ROC} &
\textbf{PR} \\
\hline

\multirow{2}{*}{SPID}
& No  & 0.6744 & 0.3000 & 0.3000 & 0.3000 & 0.5303 & 0.3212 \\
& Yes & 0.6047 & 0.3333 & 0.7000 & 0.4516 & 0.6970 & 0.3714 \\
\hline

\multirow{2}{*}{CWRU}
& No  & 0.7759 & 0.0000 & 0.0000 & 0.0000 & 0.6395 & 0.3744 \\
& Yes & 0.2069 & 0.2069 & 1.0000 & 0.3429 & 0.6359 & 0.4414 \\
\hline

\multirow{2}{*}{EFDD}
& No  & 0.5645 & 0.1852 & 0.5000 & 0.2703 & 0.5712 & 0.2484 \\
& Yes & 0.1935 & 0.1667 & 1.0000 & 0.2857 & 0.5808 & 0.2547 \\
\hline

\multirow{2}{*}{SFDD}
& No  & 0.4146 & 0.4146 & 1.0000 & 0.5862 & 0.5222 & 0.4301 \\
& Yes & 0.4146 & 0.4146 & 1.0000 & 0.5862 & 0.5093 & 0.4267 \\
\hline

\end{tabular}
\end{table}
The effect of QSMOTE was evaluated by training QTN-RQNN with and
without class balancing under the same dataset-specific preprocessing
and test partitions (Table~\ref{tab:qsmote_all}). Its impact is
dataset dependent. On SPID, QSMOTE improves recall
($0.3000\!\rightarrow\!0.7000$), F1-score
($0.3000\!\rightarrow\!0.4516$), and both ROC- and PR-AUC, despite
lower accuracy. On CWRUBD, it changes the model from zero fault
detection to a recall of $1.0000$ and F1-score of $0.3429$, again at
the cost of accuracy. EFDD shows a similar trade-off, with recall
increasing from $0.5000$ to $1.0000$ and modest improvements in
F1-score, ROC-AUC, and PR-AUC. In contrast, QSMOTE provides no benefit on SFDD: the
threshold-dependent metrics remain unchanged, while ROC- and PR-AUC
decrease slightly. QSMOTE primarily improves minority-class
sensitivity on SPID, CWRUBD, and EFDD rather than uniformly improving
all metrics, emphasizing the importance of considering recall, F1-score,
ROC-AUC, and PR-AUC alongside accuracy for imbalanced fault diagnosis.

% ============================================================
\subsection{Effect of Recurrent Timesteps}
\label{subsec:timestep_ablation}
% ============================================================

\begin{table}[!t]
\centering
\caption{Effect of recurrent timesteps $T$ on QTN-RQNN across the
four fault-diagnosis datasets.}
\label{tab:timestep_all}
\footnotesize
\setlength{\tabcolsep}{3.0pt}
\renewcommand{\arraystretch}{1.03}
\begin{tabular}{lcccccc}
\hline
\textbf{Dataset} &
$\mathbf{T}$ &
\textbf{Acc.} &
\textbf{Rec.} &
\textbf{F1} &
\textbf{ROC} &
\textbf{2Q} \\
\hline

\multirow{4}{*}{SPID}
& 1 & 0.2326 & 1.0000 & 0.3774 & 0.4545 & 6  \\
& 2 & 0.5581 & 0.7000 & 0.4242 & 0.6545 & 12 \\
& 4 & 0.6047 & 0.7000 & 0.4516 & 0.6970 & 24 \\
& 8 & 0.4186 & 0.9000 & 0.4186 & 0.5848 & 48 \\
\hline

\multirow{4}{*}{CWRUBD}
& 1 & 0.2069 & 1.0000 & 0.3429 & 0.5036 & 6  \\
& 2 & 0.8793 & 0.9167 & 0.7586 & 0.9475 & 12 \\
& 4 & 0.2069 & 1.0000 & 0.3429 & 0.6359 & 24 \\
& 8 & 0.6552 & 1.0000 & 0.5455 & 0.9420 & 48 \\
\hline

\multirow{4}{*}{EFDD}
& 1 & 0.1613 & 1.0000 & 0.2778 & 0.6423 & 6  \\
& 2 & 0.1613 & 1.0000 & 0.2778 & 0.5000 & 12 \\
& 4 & 0.1935 & 1.0000 & 0.2857 & 0.5808 & 24 \\
& 8 & 0.2581 & 0.7000 & 0.2333 & 0.5615 & 48 \\
\hline

\multirow{4}{*}{SFDD}
& 1 & 0.4146 & 1.0000 & 0.5862 & 0.5091 & 6  \\
& 2 & 0.4146 & 1.0000 & 0.5862 & 0.5334 & 12 \\
& 4 & 0.4146 & 1.0000 & 0.5862 & 0.5093 & 24 \\
& 8 & 0.4146 & 1.0000 & 0.5862 & 0.4985 & 48 \\
\hline

\end{tabular}
\end{table}

\begin{figure*}[!t]
    \centering

    \begin{subfigure}[b]{0.235\textwidth}
        \centering
        \includegraphics[width=\linewidth]
        {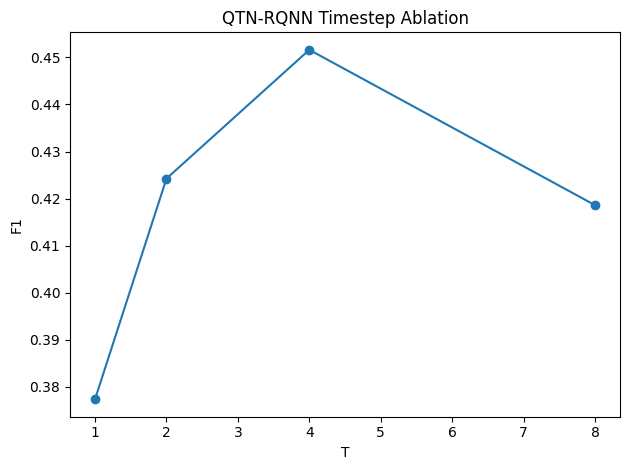}
        \caption{SPID.}
        \label{fig:spid_timestep}
    \end{subfigure}
    \hfill
    \begin{subfigure}[b]{0.235\textwidth}
        \centering
        \includegraphics[width=\linewidth]
        {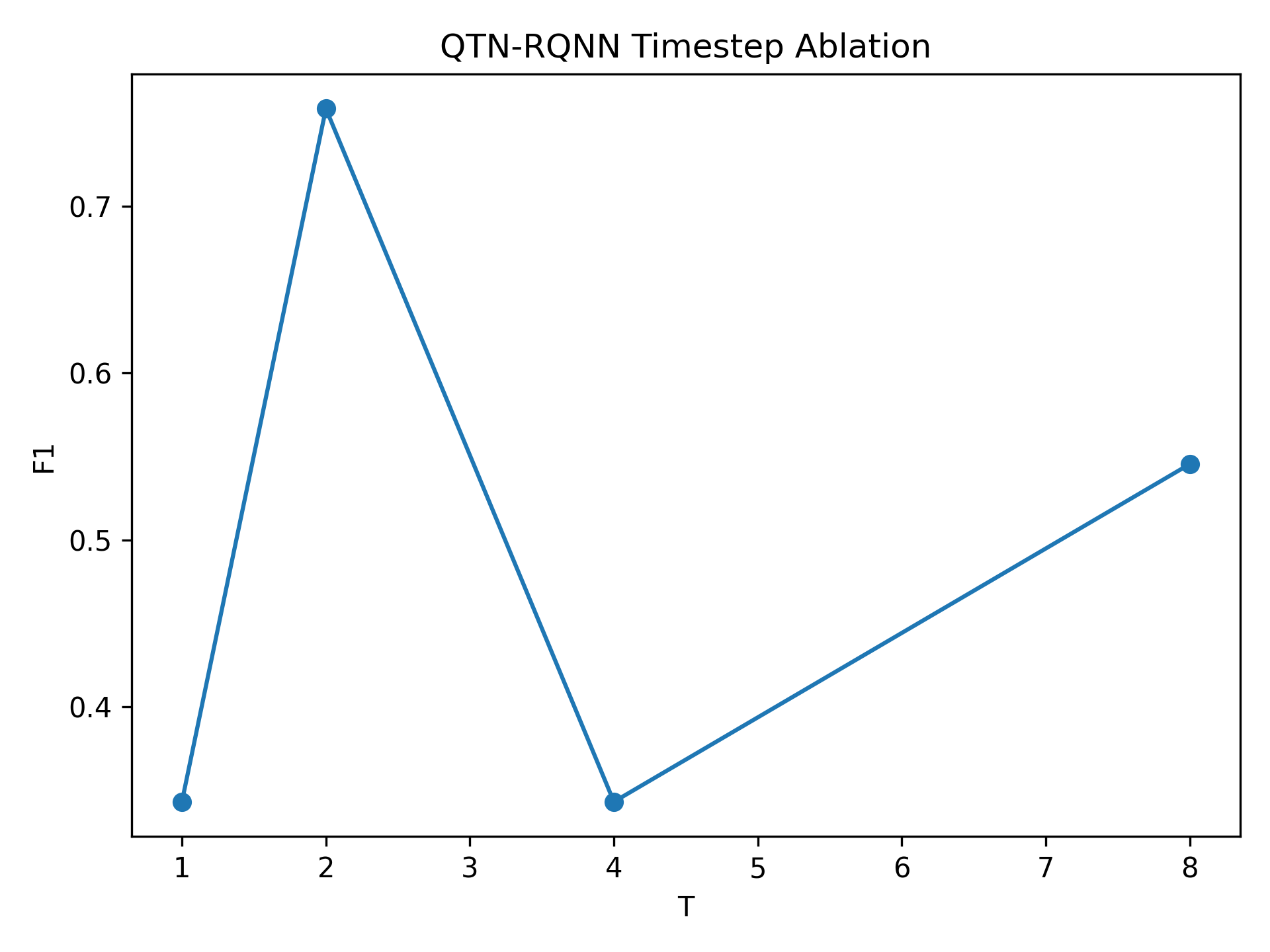}
        \caption{CWRUBD.}
        \label{fig:cwru_t_ablation}
    \end{subfigure}
    \hfill
    \begin{subfigure}[b]{0.235\textwidth}
        \centering
        \includegraphics[width=\linewidth]
        {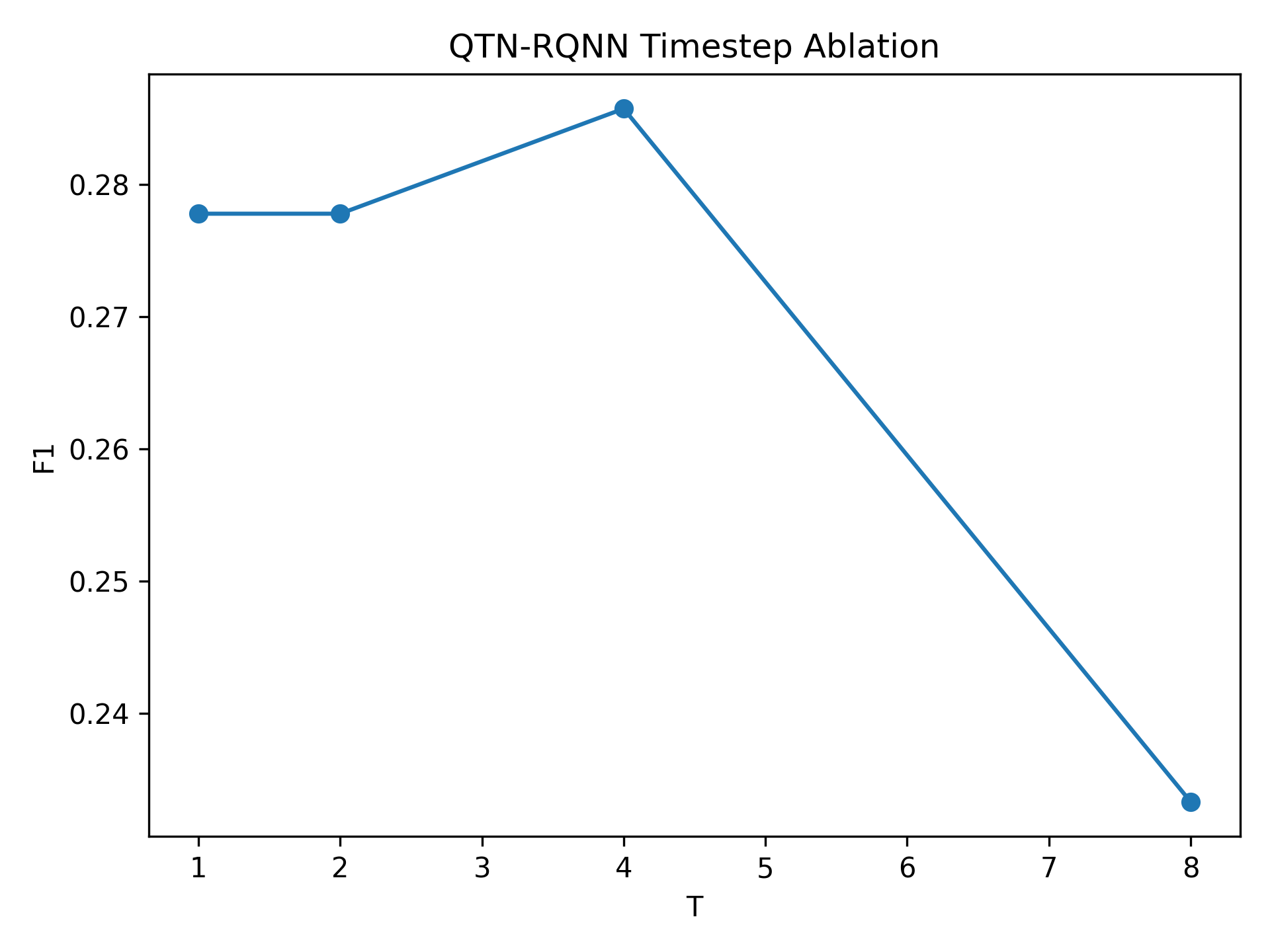}
        \caption{EFDD.}
        \label{fig:efdd_t_ablation}
    \end{subfigure}
    \hfill
    \begin{subfigure}[b]{0.235\textwidth}
        \centering
        \includegraphics[width=\linewidth]
        {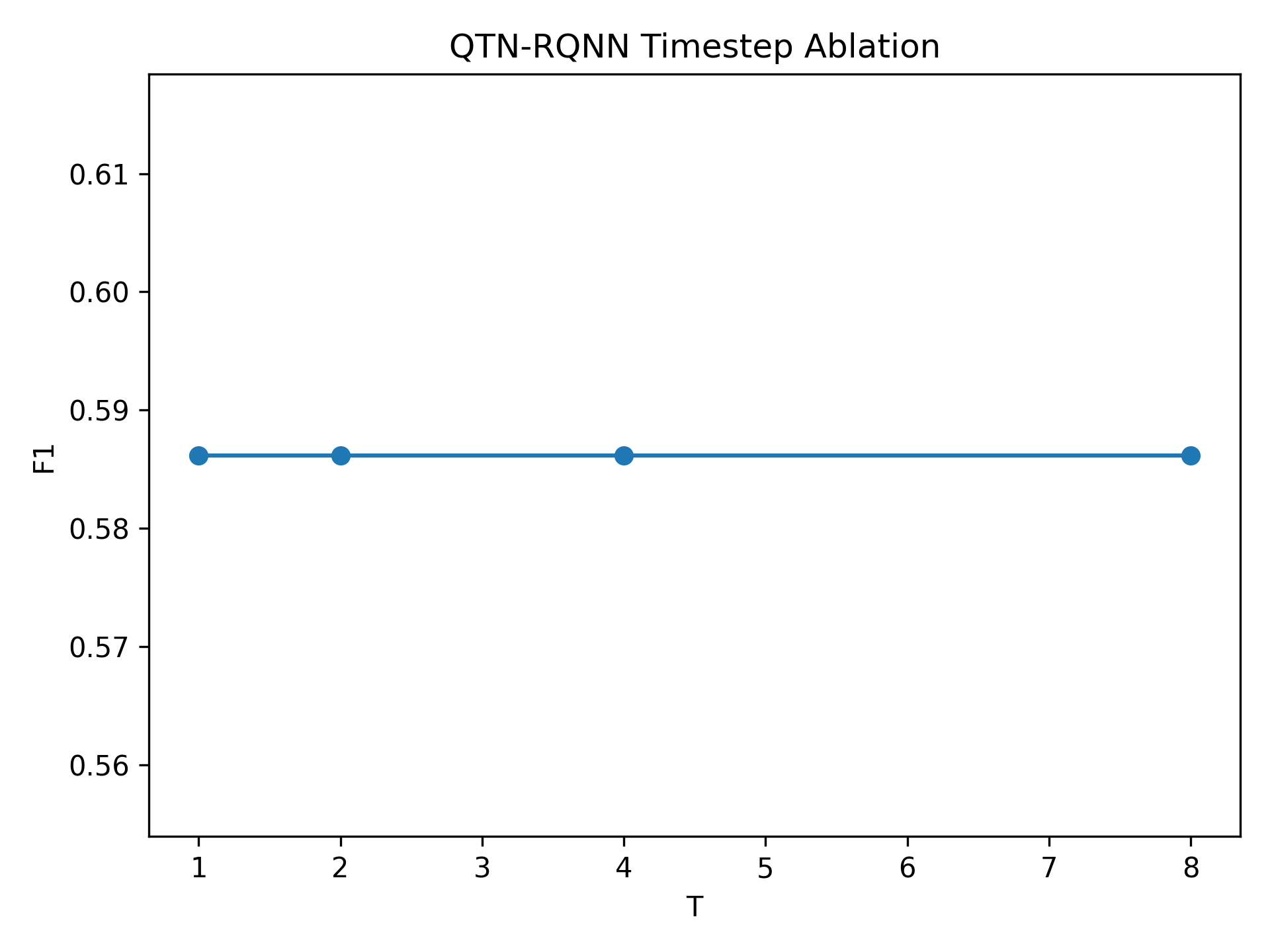}
        \caption{SFDD.}
        \label{fig:sfdd_t_ablation}
    \end{subfigure}

    \caption{Effect of the number of recurrent timesteps $T$ on
    QTN-RQNN for (a) SPID, (b) CWRUBD, (c) EFDD, and (d) SFDD.}
    \label{fig:timestep_all}
\end{figure*}
The effect of recurrent timesteps was evaluated for
$T\in\{1,2,4,8\}$, with results summarized in
Table~\ref{tab:timestep_all} and Fig.~\ref{fig:timestep_all}.
Performance is strongly dataset dependent: SPID improves up to
$T=4$, while CWRUBD performs best at $T=2$ (F1-score $0.7586$,
ROC-AUC $0.9475$), requiring only 12 two-qubit gates compared with
24 at the default $T=4$. EFDD shows comparatively modest variation,
whereas SFDD exhibits unchanged threshold-dependent performance and
only small ROC-AUC differences across timesteps. Increasing $T$ does not yield monotonic predictive gains,
while the two-qubit gate count grows linearly from 6 at $T=1$ to 48
at $T=8$. Thus, the recurrent timestep should be selected as a
dataset-dependent performance-resource trade-off rather than assuming
that longer recurrence necessarily improves predictive capability.
% ============================================================
\subsection{Effect of QTN Depth}
\label{subsec:depth_ablation}
% ============================================================

\begin{table}[!t]
\centering
\caption{Effect of QTN depth $L$ on QTN-RQNN across the four
fault-diagnosis datasets.}
\label{tab:depth_all}
\footnotesize
\setlength{\tabcolsep}{2.4pt}
\renewcommand{\arraystretch}{1.03}
\begin{tabular}{lccccccc}
\hline
\textbf{Dataset} &
$\mathbf{L}$ &
\textbf{Acc.} &
\textbf{Rec.} &
\textbf{F1} &
\textbf{ROC} &
$\mathbf{N_{\theta}}$ &
\textbf{2Q} \\
\hline

\multirow{3}{*}{SPID}
& 1 & 0.6279 & 0.8000 & 0.5000 & 0.6848 & 11 & 12 \\
& 2 & 0.6047 & 0.7000 & 0.4516 & 0.6970 & 22 & 24 \\
& 3 & 0.6977 & 0.6000 & 0.4800 & 0.7667 & 33 & 36 \\
\hline

\multirow{3}{*}{CWRUBD}
& 1 & 0.2069 & 1.0000 & 0.3429 & 0.3659 & 11 & 12 \\
& 2 & 0.2069 & 1.0000 & 0.3429 & 0.6359 & 22 & 24 \\
& 3 & 0.5690 & 1.0000 & 0.4898 & 0.9022 & 33 & 36 \\
\hline

\multirow{3}{*}{EFDD}
& 1 & 0.4355 & 0.7000 & 0.2857 & 0.5769 & 11 & 12 \\
& 2 & 0.1935 & 1.0000 & 0.2857 & 0.5808 & 22 & 24 \\
& 3 & 0.3065 & 0.8000 & 0.2712 & 0.5981 & 33 & 36 \\
\hline

\multirow{3}{*}{SFDD}
& 1 & 0.4146 & 1.0000 & 0.5862 & 0.5095 & 11 & 12 \\
& 2 & 0.4146 & 1.0000 & 0.5862 & 0.5093 & 22 & 24 \\
& 3 & 0.4146 & 1.0000 & 0.5862 & 0.4885 & 33 & 36 \\
\hline

\end{tabular}
\end{table}

\begin{figure*}[!t]
    \centering

    \begin{subfigure}[b]{0.235\textwidth}
        \centering
        \includegraphics[width=\linewidth]
        {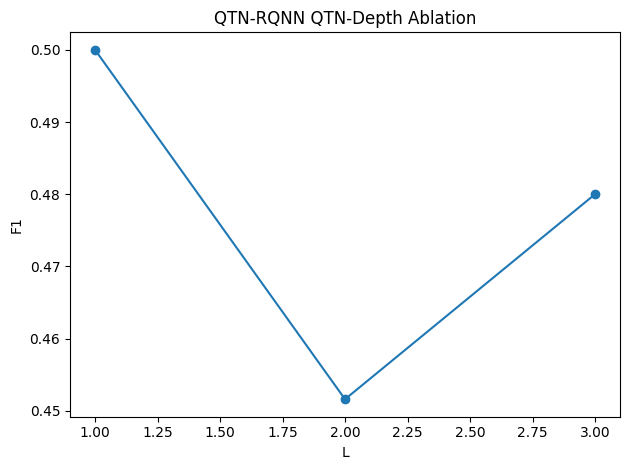}
        \caption{SPID.}
        \label{fig:spid_depth}
    \end{subfigure}
    \hfill
    \begin{subfigure}[b]{0.235\textwidth}
        \centering
        \includegraphics[width=\linewidth]
        {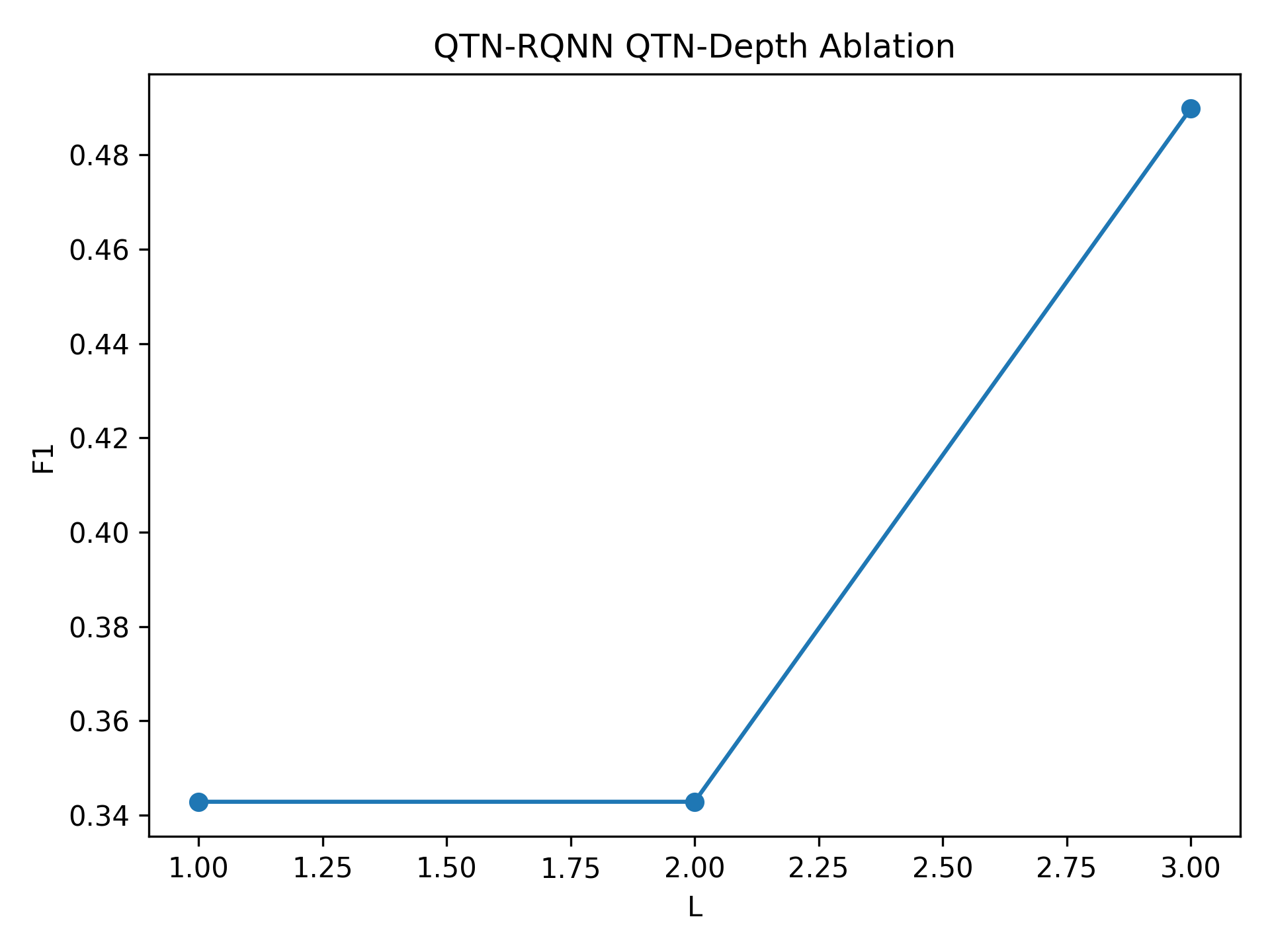}
        \caption{CWRUBD.}
        \label{fig:cwru_l_ablation}
    \end{subfigure}
    \hfill
    \begin{subfigure}[b]{0.235\textwidth}
        \centering
        \includegraphics[width=\linewidth]
        {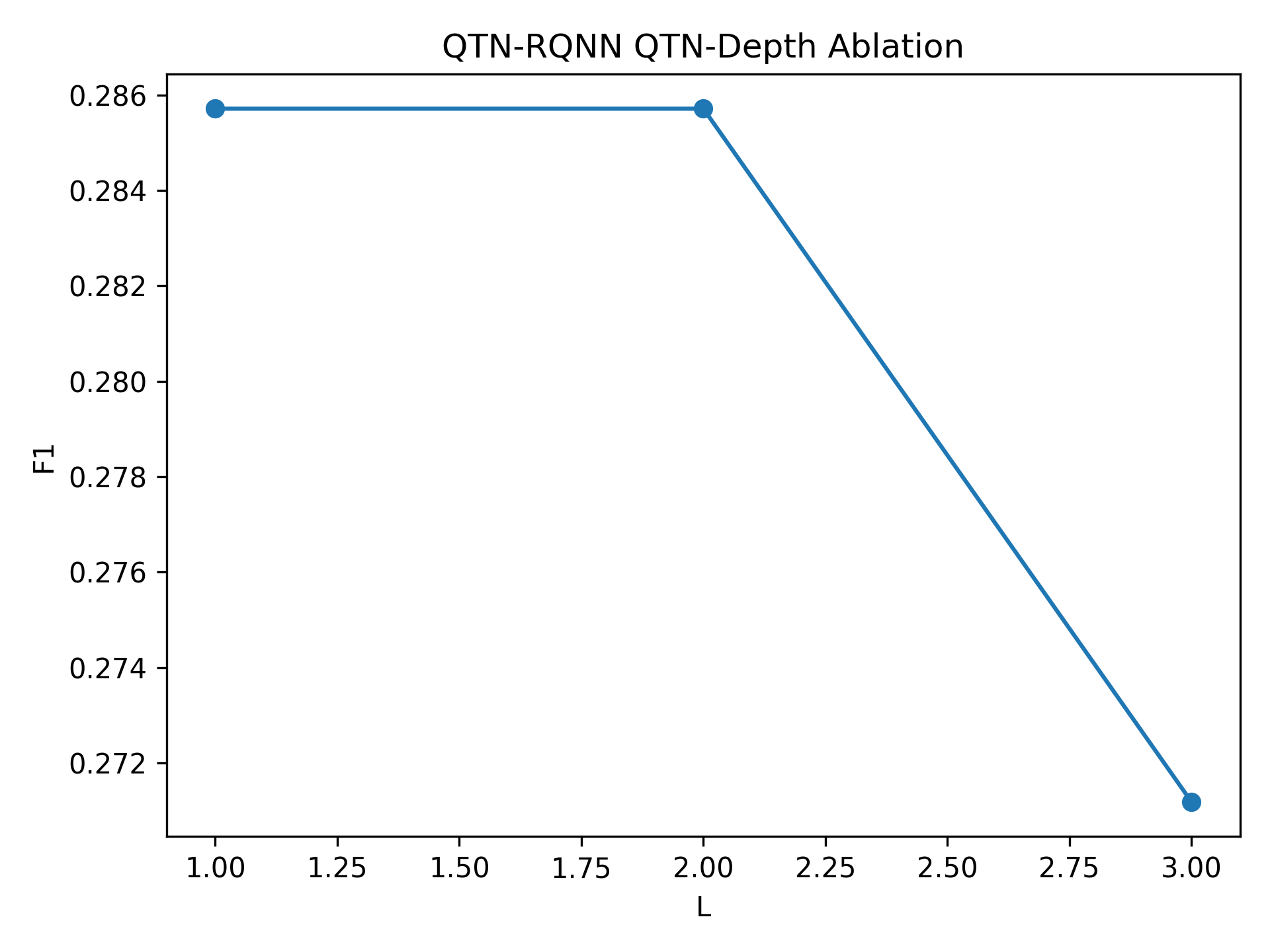}
        \caption{EFDD.}
        \label{fig:efdd_l_ablation}
    \end{subfigure}
    \hfill
    \begin{subfigure}[b]{0.235\textwidth}
        \centering
        \includegraphics[width=\linewidth]
        {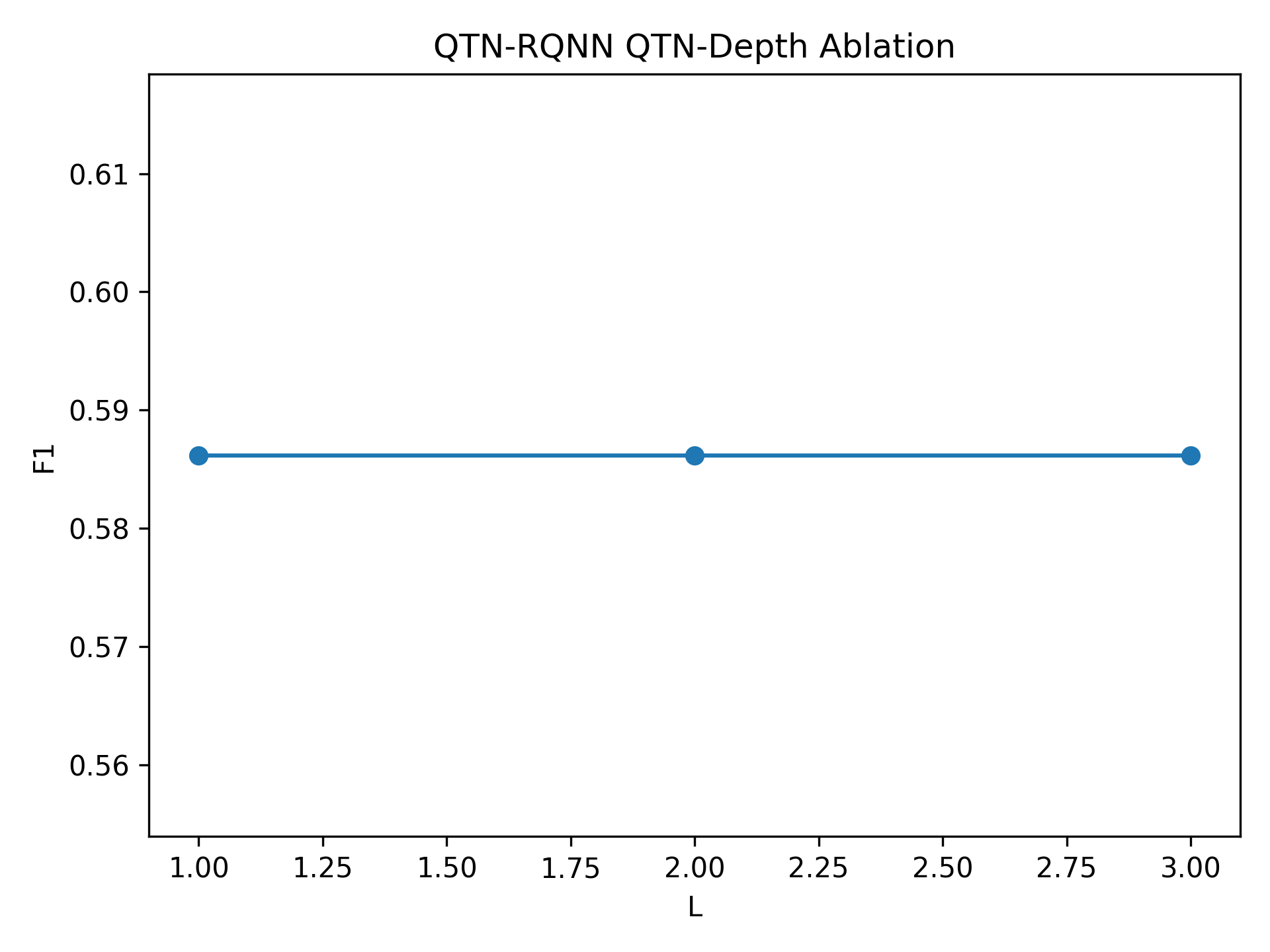}
        \caption{SFDD.}
        \label{fig:sfdd_l_ablation}
    \end{subfigure}

    \caption{Effect of QTN circuit depth $L$ on QTN-RQNN for
    (a) SPID, (b) CWRUBD, (c) EFDD, and (d) SFDD.}
    \label{fig:depth_all}
\end{figure*}
The influence of QTN depth was evaluated for $L\in\{1,2,3\}$, with
results summarized in Table~\ref{tab:depth_all} and
Fig.~\ref{fig:depth_all}. The effect is strongly dataset dependent:
SPID achieves its highest F1-score at $L=1$ but its highest accuracy
and ROC-AUC at $L=3$, while CWRUBD benefits most from $L=3$, reaching
an F1-score of $0.4898$ and ROC-AUC of $0.9022$. EFDD shows only
modest variation across depths, whereas SFDD retains identical
threshold-dependent performance with only minor ROC-AUC changes. Increasing depth nevertheless raises the resource cost systematically:
from $L=1$ to $L=3$, the number of trainable quantum parameters
increases from 11 to 33 and the two-qubit-gate count from 12 to 36
per sample. Thus, additional QTN depth does not provide monotonic
predictive gains and should be selected according to the
dataset-specific performance-resource trade-off.
% ============================================================
\subsection{Multi-Seed Robustness and Optimization Behavior}
\label{subsec:multiseed}
% ============================================================

\begin{table*}[!t]
\centering
\caption{Multi-seed robustness of QTN-RQNN across the four
fault-diagnosis datasets. Values are reported as mean $\pm$ standard
deviation over five optimizer seeds.}
\label{tab:multiseed_all}
\resizebox{\textwidth}{!}{%
\begin{tabular}{lcccccc}
\hline
\textbf{Dataset} &
\textbf{Accuracy} &
\textbf{Precision} &
\textbf{Recall} &
\textbf{F1} &
\textbf{ROC-AUC} &
\textbf{PR-AUC} \\
\hline
SPID &
$0.5163 \pm 0.1484$ &
$0.3188 \pm 0.0476$ &
$0.8600 \pm 0.1342$ &
$0.4589 \pm 0.0399$ &
$0.6564 \pm 0.0907$ &
$0.3890 \pm 0.0957$ \\

CWRUBD &
$0.6931 \pm 0.1910$ &
$0.4478 \pm 0.2024$ &
$0.7500 \pm 0.2500$ &
$0.5260 \pm 0.1728$ &
$0.8199 \pm 0.1066$ &
$0.6518 \pm 0.1829$ \\

EFDD &
$0.3742 \pm 0.0970$ &
$0.1673 \pm 0.0113$ &
$0.7200 \pm 0.1304$ &
$0.2703 \pm 0.0170$ &
$0.5658 \pm 0.0665$ &
$0.2713 \pm 0.0584$ \\

SFDD &
$0.4146 \pm 0.0000$ &
$0.4146 \pm 0.0000$ &
$1.0000 \pm 0.0000$ &
$0.5862 \pm 0.0000$ &
$0.5020 \pm 0.0103$ &
$0.4231 \pm 0.0047$ \\
\hline
\end{tabular}%
}
\end{table*}

\begin{figure*}[!t]
    \centering

    \begin{subfigure}[b]{0.235\textwidth}
        \centering
        \includegraphics[width=\linewidth]
        {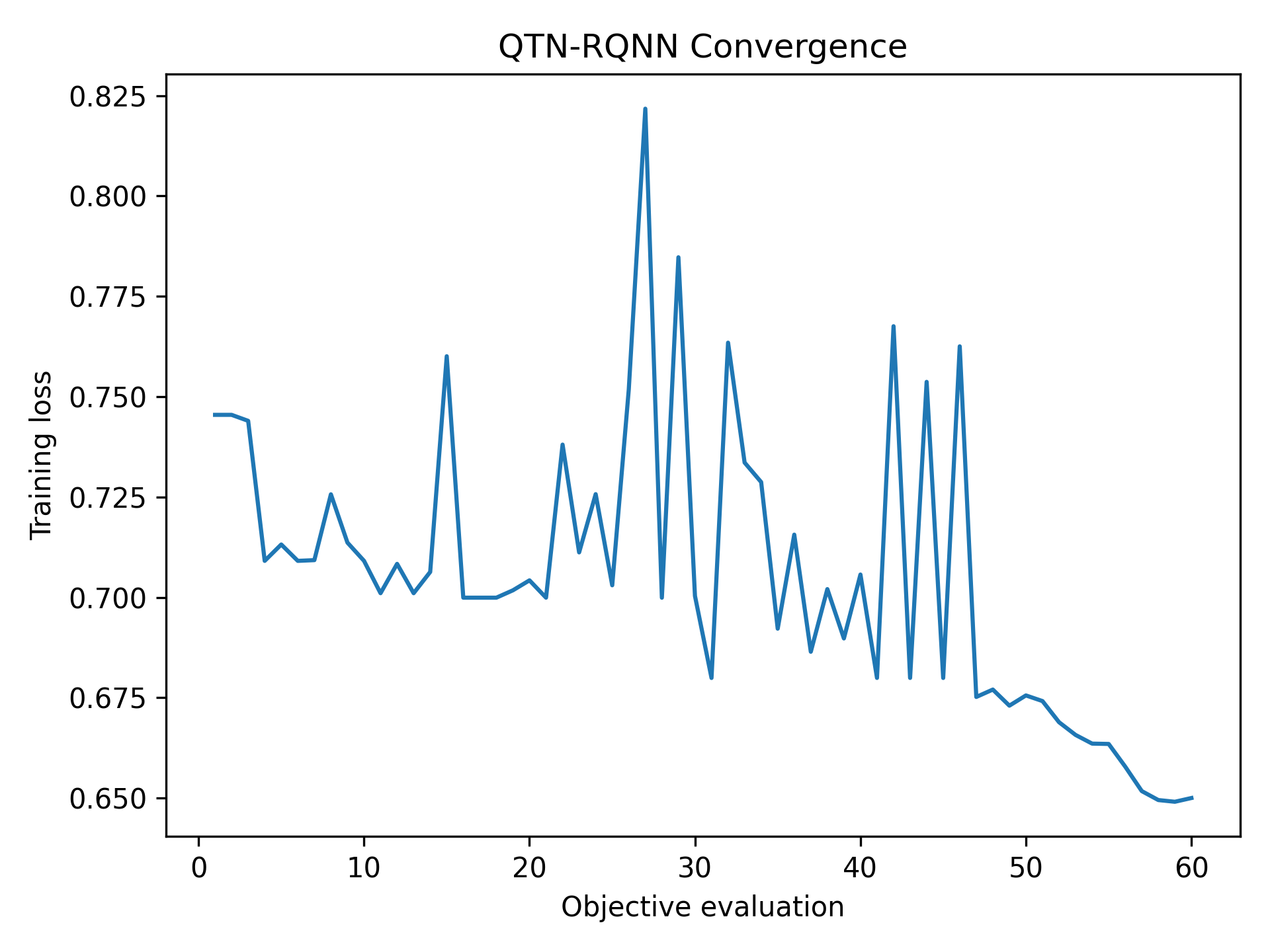}
        \caption{SPID.}
        \label{fig:spid_convergence}
    \end{subfigure}
    \hfill
    \begin{subfigure}[b]{0.235\textwidth}
        \centering
        \includegraphics[width=\linewidth]
        {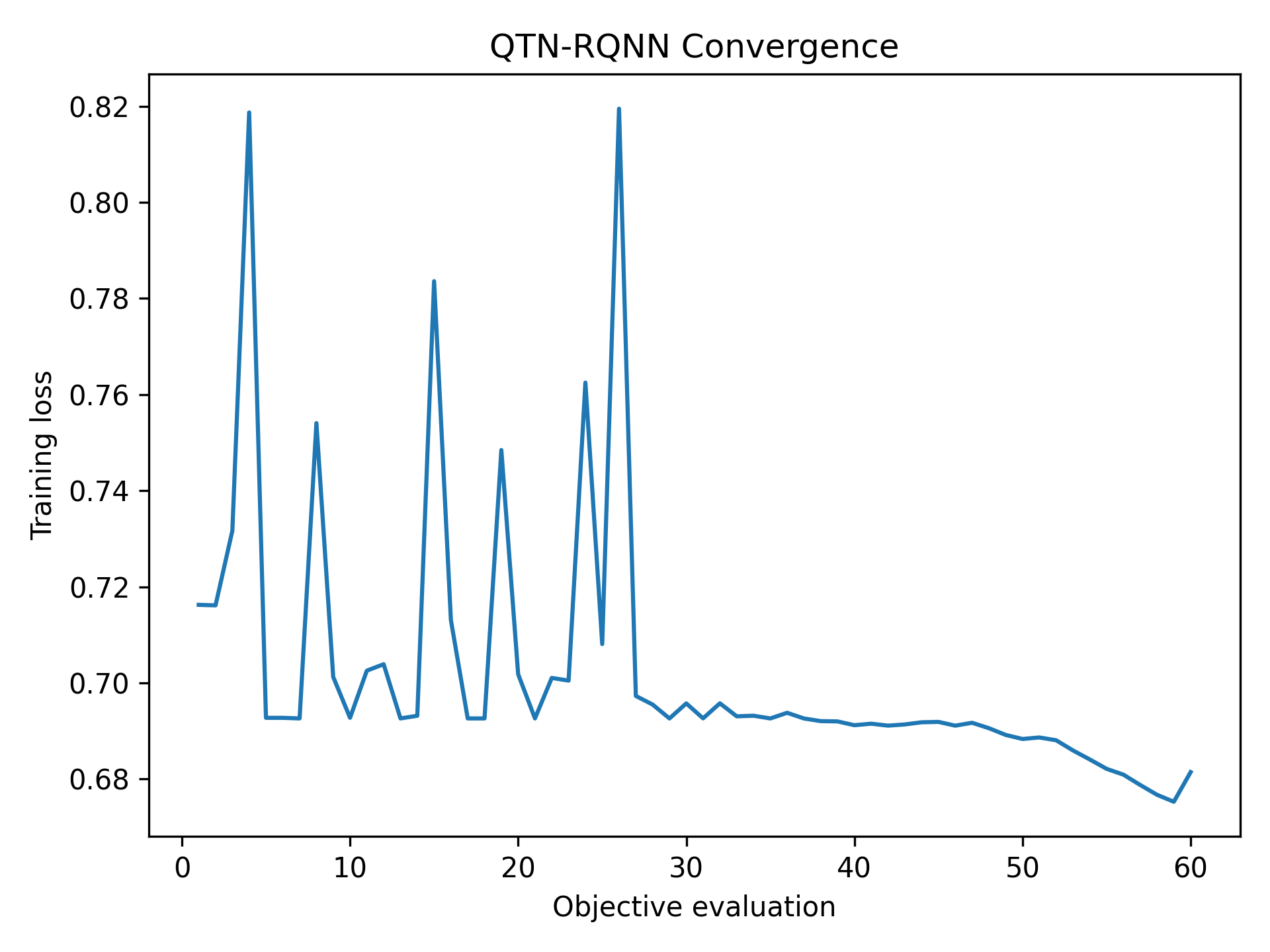}
        \caption{CWRUBD.}
        \label{fig:cwru_convergence}
    \end{subfigure}
    \hfill
    \begin{subfigure}[b]{0.235\textwidth}
        \centering
        \includegraphics[width=\linewidth]
        {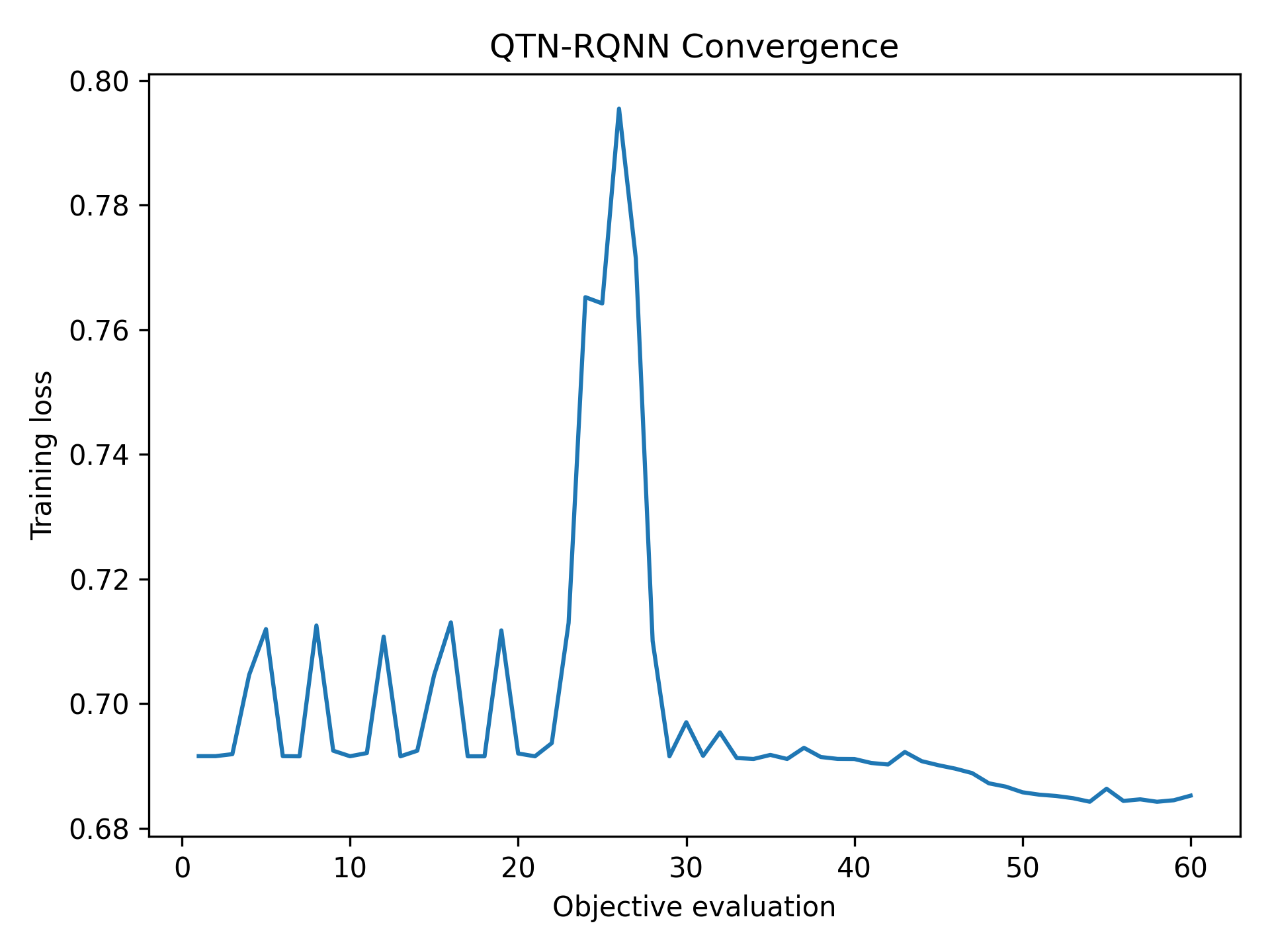}
        \caption{EFDD.}
        \label{fig:efdd_convergence}
    \end{subfigure}
    \hfill
    \begin{subfigure}[b]{0.235\textwidth}
        \centering
        \includegraphics[width=\linewidth]
        {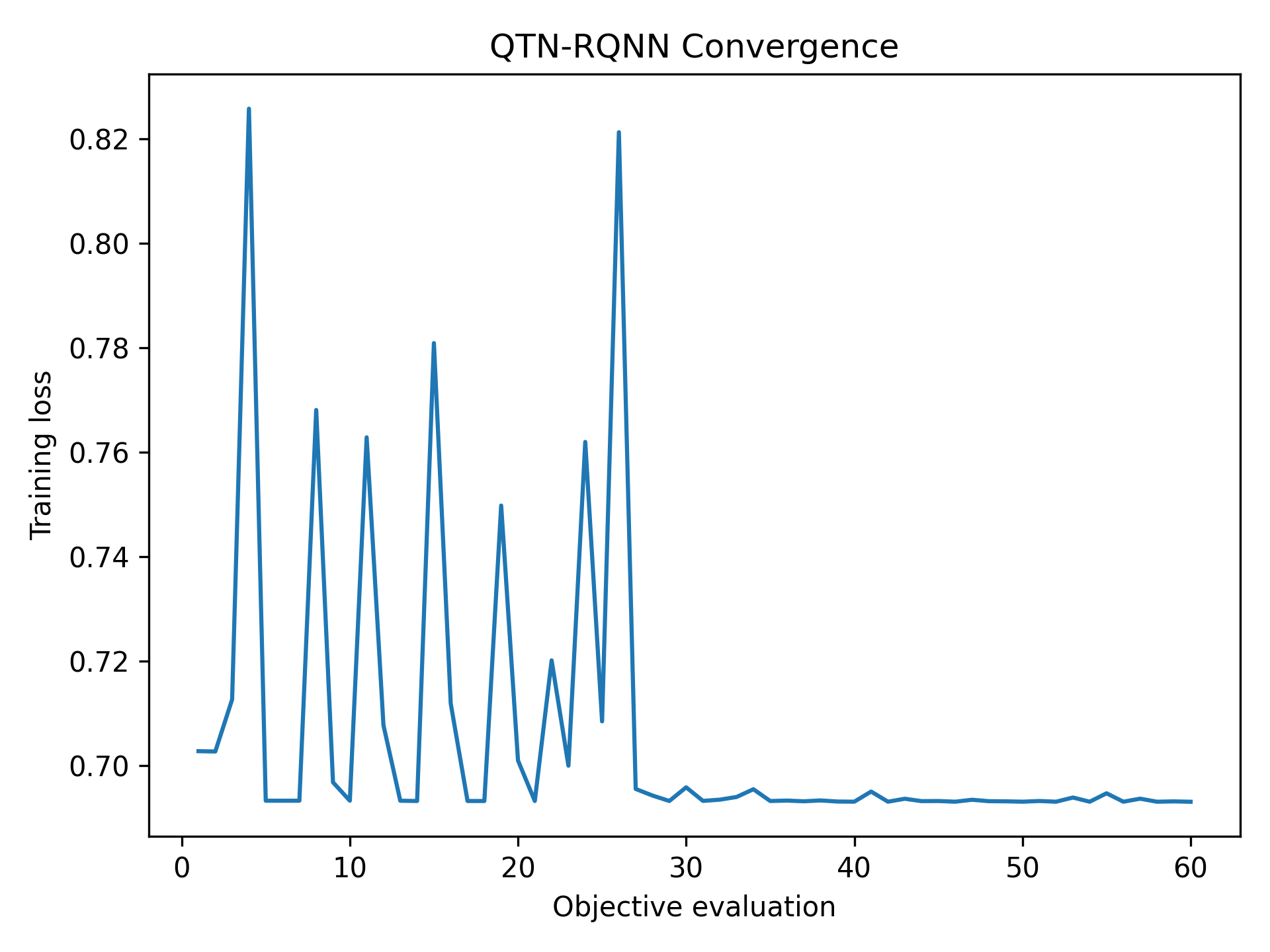}
        \caption{SFDD.}
        \label{fig:sfdd_convergence}
    \end{subfigure}

    \caption{Representative COBYLA optimization trajectories of
    QTN-RQNN for (a) SPID, (b) CWRUBD, (c) EFDD, and (d) SFDD.}
    \label{fig:convergence_all}
\end{figure*}

To assess optimization sensitivity, QTN-RQNN was evaluated using five
independent parameter-initialization seeds with a fixed data partition,
thereby separating initialization effects from split variability
(Table~\ref{tab:multiseed_all}). SPID shows relatively stable F1-score
($0.4589\pm0.0399$) despite larger accuracy variation, whereas CWRUBD
exhibits stronger seed sensitivity (F1-score $0.5260\pm0.1728$).
EFDD shows comparatively small F1-score variation
($0.2703\pm0.0170$). On SFDD, threshold-dependent metrics remain
identical across seeds; however, the mean ROC-AUC of $0.5020$ indicates
that this stability reflects weak discrimination rather than robust
predictive performance. Representative COBYLA trajectories in Fig.~\ref{fig:convergence_all}
further illustrate dataset-dependent optimization behavior, including
noticeable fluctuations for CWRUBD and EFDD before stabilization.
SFDD converges near a binary cross-entropy loss of $0.693$, consistent
with its near-random score separation. Together, these results indicate
dataset-dependent sensitivity to parameter initialization and motivate
evaluation across multiple initializations rather than reliance on a
single optimization run.

% ============================================================
\subsection{Quantum-Resource Analysis}
\label{subsec:resource_analysis}
% ============================================================

\begin{table*}[!t]
\centering
\caption{Quantum-resource comparison of the evaluated architectures.
The reported counts are identical across SPID, CWRUBD, EFDD, and SFDD.}
\label{tab:resource_all}
\begin{tabular}{lcccccc}
\hline
\textbf{Model} &
\textbf{Qubits} &
\textbf{Quantum Params.} &
\textbf{Total Params.} &
\textbf{1Q Gates/Sample} &
\textbf{2Q Gates/Sample} &
\textbf{Cell Repetitions} \\
\hline
QNN      & 4 & 16 & 19 & 24 & 8  & 1 \\
QTN-QNN  & 4 & 22 & 25 & 24 & 6  & 1 \\
RQNN     & 4 & 16 & 19 & 72 & 32 & 4 \\
QTN-RQNN & 4 & 22 & 25 & 72 & 24 & 4 \\
\hline
\end{tabular}
\end{table*}

Beyond predictive performance, the architectures were compared in
terms of quantum-resource requirements
(Table~\ref{tab:resource_all}). Since identical circuit configurations
are used across datasets, these counts are dataset independent and
refer to one forward evaluation per sample. All architectures require
four qubits. The QTN-based models use 22 quantum (25 total) trainable
parameters, compared with 16 quantum (19 total) parameters for their
conventional counterparts, reflecting a trade-off between increased
parameterization and reduced two-qubit connectivity. Specifically, QTN-QNN reduces the two-qubit-gate count from 8 to 6
relative to QNN, while QTN-RQNN reduces it from 32 to 24 relative to
RQNN, corresponding to a $25\%$ reduction in both cases. Recurrence
increases the one-qubit-gate count from 24 to 72 and the number of
cell executions from one to four; therefore, resource comparisons are
most meaningful within the feedforward (QNN/QTN-QNN) and recurrent
(RQNN/QTN-RQNN) pairs. Thus, QTN-RQNN preserves the four-qubit
requirement and recurrent processing structure of RQNN while reducing
its two-qubit interactions, providing a resource-conscious circuit
construction despite its larger number of trainable parameters.

\subsection{Discussion}

The experiments demonstrate that QTN-RQNN exhibits strongly
dataset-dependent behavior rather than a universal predictive
advantage over simpler quantum architectures. QTN structuring improves
the recurrent model on SPID, whereas CWRUBD is particularly sensitive
to recurrent segmentation and circuit depth. On EFDD, QTN-RQNN favors
fault sensitivity at the expense of false positives, while the
near-random ROC-AUC and all-positive operating point on SFDD expose
the limitations of the present compact representation. The ablation studies further show that increasing recurrent timesteps
or QTN depth does not produce monotonic improvements, while the effect
of QSMOTE and parameter initialization is also dataset dependent.
In contrast, the QTN-based circuits provide a consistent structural
resource benefit: QTN-QNN and QTN-RQNN require $25\%$ fewer two-qubit
gates than their corresponding conventional architectures without
increasing the qubit count, although with additional trainable
parameters. These results motivate evaluating QTN-RQNN through
the joint trade-off between predictive performance and quantum-resource
requirements rather than architectural complexity alone.

\section{Conclusion}\label{SecV}

This work introduced QTN-RQNN, a recurrent quantum learning
architecture combining sequential feature encoding, propagated quantum
memory, and tensor-network-inspired nearest-neighbor interactions for
fault classification. Evaluation on SPID, CWRUBD, EFDD, and SFDD
against QNN, QTN-QNN, and RQNN baselines demonstrates strongly
dataset-dependent predictive behavior, with the ablation and
multi-seed studies further showing sensitivity to recurrent
segmentation, circuit depth, class balancing, and initialization. The principal architecture-level benefit is a consistent reduction in
two-qubit interactions. Under the four-qubit configuration, QTN-RQNN
reduces the two-qubit-gate count from 32 to 24 relative to RQNN, while
QTN-QNN reduces it from 8 to 6 relative to QNN, corresponding to a
$25\%$ reduction in both cases. This benefit is obtained without
increasing the qubit requirement, although with additional trainable
parameters. Thus, QTN-RQNN should be viewed as a resource-conscious
recurrent architecture rather than a universally superior classifier.
Future work will investigate noise-aware and hardware-based validation,
transpiled resource costs, alternative QTN topologies, and adaptive
encoding and circuit-depth strategies.

\bibliographystyle{IEEEtran}
\bibliography{IEEE}

\section*{Appendix: Supplementary Methodological Details}
\label{supp:methodology}

This Supplement provides implementation and derivation details omitted
from the main manuscript for compactness. The notation follows
Section~III of the main text.

\subsection{Dataset-Specific Preprocessing}

All preprocessing transformations are determined using the training
partition and subsequently applied to validation and test data to avoid
information leakage. Each dataset is transformed into an
eight-dimensional representation before recurrent quantum processing. For datasets whose processed feature dimension exceeds eight,
dimensionality reduction is applied using the training partition.
For an input matrix $\mathbf{X}\in\mathbb{R}^{N\times d}$, the
centered covariance matrix is
\begin{equation}
\mathbf{C}
=
\frac{1}{N-1}
(\mathbf{X}-\boldsymbol{\mu})^{\top}
(\mathbf{X}-\boldsymbol{\mu}),
\end{equation}
and the retained components are determined from
\begin{equation}
\mathbf{C}\mathbf{v}_{j}
=
\lambda_j\mathbf{v}_{j}.
\end{equation}
The transformed representation is
\begin{equation}
\mathbf{x}_{\mathrm{PCA}}
=
\mathbf{V}_{8}^{\top}
(\mathbf{x}-\boldsymbol{\mu}),
\end{equation}
where $\mathbf{V}_{8}$ contains the eight retained eigenvectors. Each feature is subsequently scaled using training-set extrema,
\begin{equation}
x_j'
=
\frac{x_j-x_j^{\min}}
{x_j^{\max}-x_j^{\min}},
\end{equation}
yielding the $[0,1]$ interval required by the implemented angle
encoder. Dataset-specific preprocessing choices and feature extraction
settings are reported with the experimental setup.

\subsection{QSMOTE-Based Training-Class Balancing}

QSMOTE is applied only to the training partition. Validation and test
samples remain unchanged throughout class balancing. The implementation
first partitions the training samples into clusters using $K$-means.
For a selected sample $\mathbf{x}$ and its assigned centroid
$\mathbf{c}$, define
\begin{equation}
Z
=
\|\mathbf{x}\|_2^2+\|\mathbf{c}\|_2^2.
\end{equation}

A compact swap-test-based routine estimates a measurement probability
$p_0$. The implemented overlap-related quantity is then evaluated as \cite{Patel2026NoisyQSMOTE}
\begin{equation}
s
=
\frac{(3-4p_0)Z}
{2\|\mathbf{x}\|_2\|\mathbf{c}\|_2},
\end{equation}
with numerical clipping to $[-1,1]$. The corresponding angular
quantity is
\begin{equation}
\alpha
=
\arccos
\left[
\operatorname{clip}(s,-1,1)
\right].
\end{equation}

A displacement magnitude is sampled according to
\begin{equation}
r
\sim
\mathcal{U}
\left(
0,\frac{\alpha}{s_f}
\right),
\end{equation}
where $s_f$ is the split factor. The synthetic observation is generated
along the normalized direction between the selected sample and its
cluster centroid,
\begin{equation}
\mathbf{x}_{\mathrm{syn}}
=
\mathbf{x}
+
r
\frac{\mathbf{c}-\mathbf{x}}
{\|\mathbf{c}-\mathbf{x}\|_2}.
\end{equation}

Synthetic observations are generated until the target training-class
size is reached. No synthetic observations are generated for validation
or test partitions.

\subsection{Construction of the Recurrent Input State}

For the default configuration, the eight-dimensional normalized vector
is partitioned as
\begin{equation}
\mathbf{x}
=
[
x_1,x_2,
x_3,x_4,
x_5,x_6,
x_7,x_8
],
\end{equation}
with
\begin{align}
\mathbf{x}_{1}&=(x_1,x_2),\\
\mathbf{x}_{2}&=(x_3,x_4),\\
\mathbf{x}_{3}&=(x_5,x_6),\\
\mathbf{x}_{4}&=(x_7,x_8).
\end{align}

At timestep $t$, the input state is
\begin{equation}
|\phi_t\rangle
=
R_Y(\pi x_{t,1})|0\rangle
\otimes
R_Y(\pi x_{t,2})|0\rangle.
\end{equation}
Its density operator is
\begin{equation}
\rho_{I,t}
=
|\phi_t\rangle\langle\phi_t|.
\end{equation}

Given the memory state $\rho_{M,t-1}$, the complete four-qubit state
passed to the recurrent cell is constructed as
\begin{equation}
\rho_{IM,t}
=
\rho_{I,t}\otimes\rho_{M,t-1}.
\end{equation}

\subsection{Detailed QTN Parameterization}

For each layer $\ell$, the local trainable block contains
\begin{equation}
\boldsymbol{\theta}_{\ell}^{\mathrm{loc}}
=
\{
\theta_{\ell,q}^{y},
\theta_{\ell,q}^{z}
\}_{q=0}^{3},
\end{equation}
giving eight local parameters. The entangling parameter set is
\begin{equation}
\boldsymbol{\gamma}_{\ell}
=
\{
\gamma_{\ell,01},
\gamma_{\ell,23},
\gamma_{\ell,12}
\},
\end{equation}
giving three additional parameters. Hence,
\begin{equation}
\boldsymbol{\theta}_{\ell}
=
[
\boldsymbol{\theta}_{\ell}^{\mathrm{loc}},
\boldsymbol{\gamma}_{\ell}
]
\in\mathbb{R}^{11}.
\end{equation}

For $L$ layers,
\begin{equation}
\boldsymbol{\theta}
=
[
\boldsymbol{\theta}_{1},
\ldots,
\boldsymbol{\theta}_{L}
]
\in\mathbb{R}^{11L}.
\end{equation}

The complete four-qubit recurrent unitary is
\begin{equation}
U_{\mathrm{QTN}}(\boldsymbol{\theta})
=
\prod_{\ell=L}^{1}
\left[
U_{\mathrm{odd}}^{(\ell)}
U_{\mathrm{even}}^{(\ell)}
U_{\mathrm{loc}}^{(\ell)}
\right].
\end{equation}

\subsection{Reduced-State Memory Update}

After applying the recurrent unitary,
\begin{equation}
\widetilde{\rho}_{IM,t}
=
U_{\mathrm{QTN}}
\rho_{IM,t}
U_{\mathrm{QTN}}^{\dagger}.
\end{equation}

The memory state is obtained by tracing over the two input qubits,
\begin{equation}
\rho_{M,t}
=
\operatorname{Tr}_{q_0,q_1}
[
\widetilde{\rho}_{IM,t}
].
\end{equation}

The reduced state is generally mixed because the input and memory
registers may become entangled before the input subsystem is removed.
The resulting recurrence can therefore be written as the
input-conditioned channel
\begin{equation}
\mathcal{E}_{\mathbf{x}_t,\boldsymbol{\theta}}
(\rho)
=
\operatorname{Tr}_{I}
\left[
U_{\mathrm{QTN}}
\left(
|\phi_t\rangle\langle\phi_t|
\otimes\rho
\right)
U_{\mathrm{QTN}}^{\dagger}
\right].
\end{equation}
Consequently,
\begin{equation}
\rho_{M,T}
=
\mathcal{E}_{\mathbf{x}_T,\boldsymbol{\theta}}
\circ\cdots\circ
\mathcal{E}_{\mathbf{x}_1,\boldsymbol{\theta}}
(\rho_{M,0}).
\end{equation}

This formulation makes explicit that the implementation propagates a
reduced two-qubit density operator rather than retaining all input
qubits coherently across the complete sequence.

\subsection{Readout and Joint Hybrid Optimization}

The final memory state produces
\begin{equation}
z_0
=
\operatorname{Tr}
[
\rho_{M,T}(Z\otimes I)
],
\qquad
z_1
=
\operatorname{Tr}
[
\rho_{M,T}(I\otimes Z)
].
\end{equation}
The complete trainable parameter vector is
\begin{equation}
\boldsymbol{\Theta}
=
[
\boldsymbol{\theta},
w_0,w_1,b
].
\end{equation}

For $L=2$,
\begin{equation}
|\boldsymbol{\Theta}|
=
22+3
=
25.
\end{equation}

The predicted probability is
\begin{equation}
\widehat{p}
=
\frac{1}
{1+\exp[-(\mathbf{w}^{\top}\mathbf{z}+b)]}.
\end{equation}

Class-weighted binary cross-entropy is used during optimization:
\begin{equation}
\mathcal{L}
=
-\frac{1}{N}
\sum_{i=1}^{N}
\left[
w_1y_i\log(\widehat{p}_i+\epsilon)
+
w_0(1-y_i)
\log(1-\widehat{p}_i+\epsilon)
\right],
\end{equation}
where $\epsilon$ is used for numerical stability.

\subsection{COBYLA Optimization and Validation Protocol}

The model is optimized using COBYLA because the implementation treats
the quantum model as a derivative-free objective. Multiple independent
parameter initializations are evaluated. For each initialization, the
quantum and classical parameters are optimized jointly. Following optimization, probabilities are evaluated on the validation
partition. The operating threshold is selected from
\begin{equation}
\mathcal{T}
=
\{0.05,0.06,\ldots,0.95\}
\end{equation}
according to
\begin{equation}
\tau^{*}
=
\arg\max_{\tau\in\mathcal{T}}
F_1(\tau).
\end{equation}
The selected parameter initialization and threshold are subsequently
used for evaluation on the untouched test partition.

\subsection{Baseline Circuit Definitions}

Three architectures are evaluated in addition to QTN-RQNN to separate
the effects of QTN structuring and recurrent processing.

\subsubsection{QNN}

The feedforward QNN operates on four qubits. The eight normalized
features are encoded using
\begin{equation}
U_{\mathrm{enc}}(\mathbf{x})
=
\prod_{q=0}^{3}
R_Z^{(q)}(\pi x_{q+4})
R_Y^{(q)}(\pi x_q).
\end{equation}
The trainable block applies local $R_Y/R_Z$ rotations followed by a
four-qubit ring of CNOT operations.

\subsubsection{QTN-QNN}

QTN-QNN uses the same feedforward encoding and readout as QNN but
replaces the ring-CNOT entangling pattern with the three-bond
nearest-neighbor QTN structure
\begin{equation}
(0,1),\;(2,3),\;(1,2).
\end{equation}
It therefore isolates the effect of QTN-inspired connectivity without
recurrence.

\subsubsection{RQNN}

RQNN uses the same input/memory partition, sequential encoding, memory
update, and classical readout as QTN-RQNN. Its recurrent quantum cell,
however, uses local $R_Y/R_Z$ rotations followed by the conventional
four-edge ring entangler. Comparison between RQNN and QTN-RQNN
therefore isolates the effect of replacing ring connectivity with the
nearest-neighbor QTN structure.

\subsection{Logical Quantum-Resource Derivation}

For the proposed QTN cell, every layer contains
\begin{equation}
N_{\mathrm{1Q}}^{\mathrm{layer}}
=
2q
\end{equation}
trainable one-qubit gates and
\begin{equation}
N_{\mathrm{2Q}}^{\mathrm{layer}}
=
q-1
\end{equation}
two-qubit gates. Including the two input-encoding rotations at every recurrent step,
the complete QTN-RQNN requires
\begin{equation}
N_{\mathrm{1Q}}^{\mathrm{QTN-RQNN}}
=
T(2+2qL)
\end{equation}
and
\begin{equation}
N_{\mathrm{2Q}}^{\mathrm{QTN-RQNN}}
=
T(q-1)L.
\end{equation}

For $q=4$, $T=4$, and $L=2$,
\begin{equation}
N_{\mathrm{1Q}}^{\mathrm{QTN-RQNN}}
=
4(2+16)
=
72,
\end{equation}
and
\begin{equation}
N_{\mathrm{2Q}}^{\mathrm{QTN-RQNN}}
=
4(3)(2)
=
24.
\end{equation}

The RQNN ring contains $q$ two-qubit interactions per layer, giving
\begin{equation}
N_{\mathrm{2Q}}^{\mathrm{RQNN}}
=
TqL
=
4(4)(2)
=
32.
\end{equation}
The relative reduction is therefore
\begin{equation}
\Delta_{\mathrm{2Q}}
=
\frac{32-24}{32}
\times100
=
25\%.
\end{equation}

For the corresponding feedforward circuits, the implemented logical
counts are six two-qubit gates for QTN-QNN and eight for QNN, again
corresponding to a $25\%$ reduction. These values represent logical circuit counts under the implemented
model and should not be interpreted as hardware-native gate counts.
Hardware-specific compilation can decompose the parameterized
operations and alter the final gate count and depth.

\subsection{Timestep Mapping Used in the Ablation Study}

The main configuration $T=4$ provides an exact partition of the
eight-dimensional representation into four two-feature segments. For
the timestep study $T\in\{1,2,4,8\}$, the implementation maps every
segment to the two-dimensional input register. Two-element segments
are retained unchanged. Single-element segments are duplicated, while
segments containing more than two values are compressed into two
values by averaging their two halves. Consequently, changing $T$ changes both the number of recurrent cell
applications and, for $T\neq4$, the deterministic mapping of the
eight-dimensional representation into the two-qubit input register.
The timestep experiment should therefore be interpreted as evaluating
the complete sequence-construction configuration rather than as a
strictly isolated recurrence-length intervention.

\subsection{Algorithmic Description of QTN-RQNN}
\label{sup:algorithm}

\begin{algorithm}[!t]
\caption{Training and Inference of QTN-RQNN}
\label{alg:qtn_rqnn}
\begin{algorithmic}[1]

\Require Training, validation, and test sets
$\mathcal{D}_{\mathrm{tr}}$, $\mathcal{D}_{\mathrm{val}}$,
$\mathcal{D}_{\mathrm{te}}$; recurrent timesteps $T$;
QTN depth $L$
\Ensure Trained parameters and test predictions

\State Fit preprocessing transformations using
$\mathcal{D}_{\mathrm{tr}}$ and transform all partitions
\State Apply QSMOTE only to $\mathcal{D}_{\mathrm{tr}}$
\State Partition each input into $T$ sequential feature segments
$\{\mathbf{x}_1,\ldots,\mathbf{x}_T\}$
\State Initialize shared QTN parameters $\boldsymbol{\theta}$ and
classical output parameters $\mathbf{w},b$

\While{optimization stopping criterion is not satisfied}
    \For{each training sample}
        \State Initialize memory
        $\rho_{M,0}=|00\rangle\langle00|$
        \For{$t=1,\ldots,T$}
            \State Encode $\mathbf{x}_t=(x_{t,1},x_{t,2})$ on the
            input qubits using
            $R_Y(\pi x_{t,1})$ and $R_Y(\pi x_{t,2})$
            \State Form
            $\rho_{IM,t}
            =|\phi_t\rangle\langle\phi_t|\otimes\rho_{M,t-1}$
            \State Apply the shared QTN cell:
            $\widetilde{\rho}_{IM,t}
            =U_{\mathrm{QTN}}(\boldsymbol{\theta})
            \rho_{IM,t}
            U_{\mathrm{QTN}}^{\dagger}(\boldsymbol{\theta})$
            \State Update memory:
            $\rho_{M,t}
            =\operatorname{Tr}_{I}
            [\widetilde{\rho}_{IM,t}]$
        \EndFor
        \State Measure the final memory:
        $\mathbf{z}
        =[\langle Z\otimes I\rangle,
        \langle I\otimes Z\rangle]^{\top}$
        \State Compute
        $\hat{p}=\sigma(\mathbf{w}^{\top}\mathbf{z}+b)$
    \EndFor
    \State Evaluate the class-weighted binary cross-entropy loss
    \State Update $\boldsymbol{\theta}$, $\mathbf{w}$, and $b$
    using COBYLA
\EndWhile

\State Select decision threshold $\tau$ using
$\mathcal{D}_{\mathrm{val}}$

\For{each test sample}
    \State Propagate its $T$ segments through the trained recurrent
    QTN cell
    \State Obtain $\hat{p}$ from the final memory measurements
    \State Predict
    $\hat{y}=\mathbb{I}[\hat{p}\geq\tau]$
\EndFor

\State \Return trained parameters and test predictions

\end{algorithmic}
\end{algorithm}

\begin{figure*}[!t]
    \centering

    \begin{subfigure}[b]{0.235\textwidth}
        \centering
        \includegraphics[width=\linewidth]
        {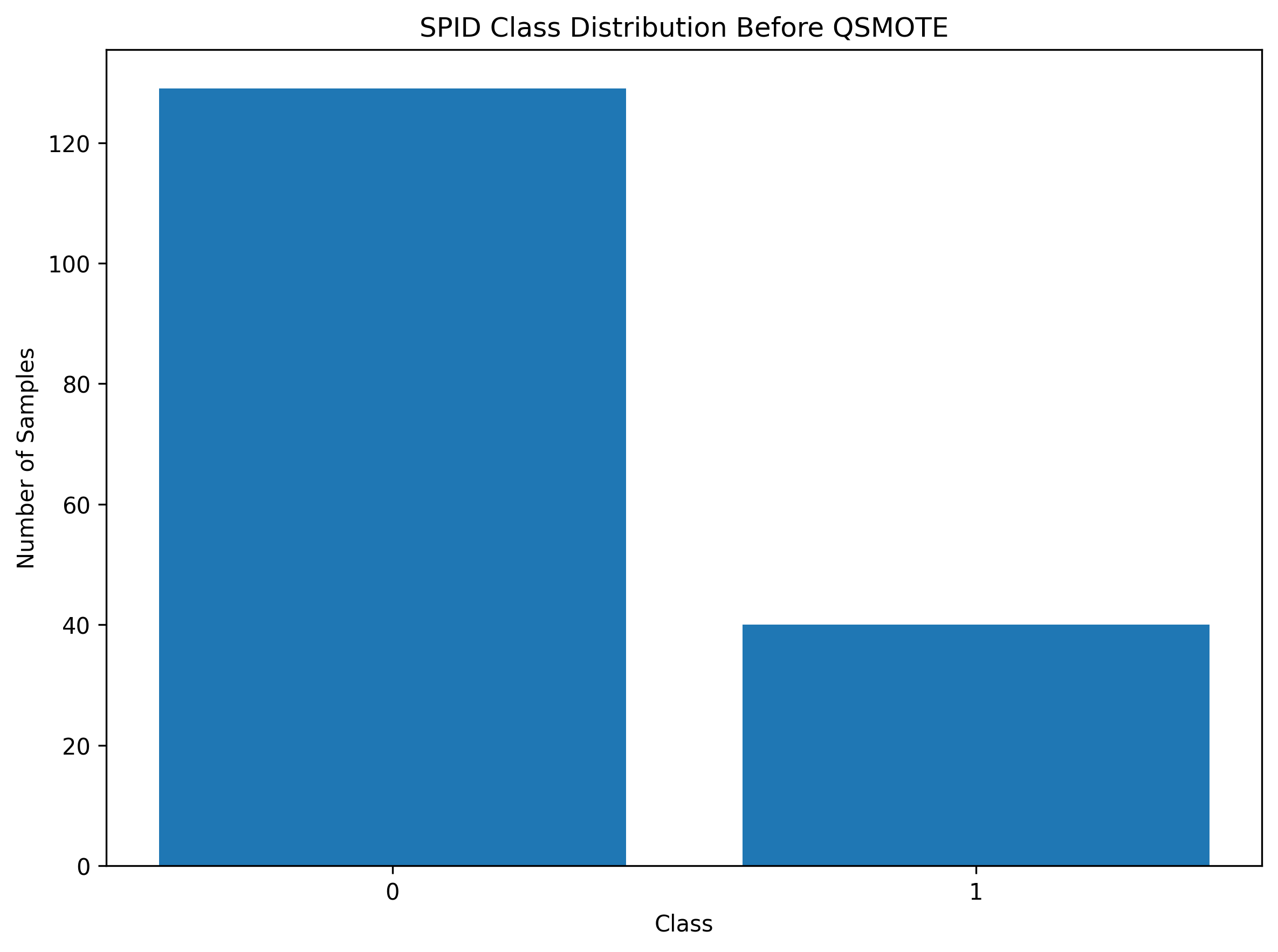}
        \caption{SPID.}
        \label{fig:sup_spid_before_qsmote}
    \end{subfigure}
    \hfill
    \begin{subfigure}[b]{0.235\textwidth}
        \centering
        \includegraphics[width=\linewidth]
        {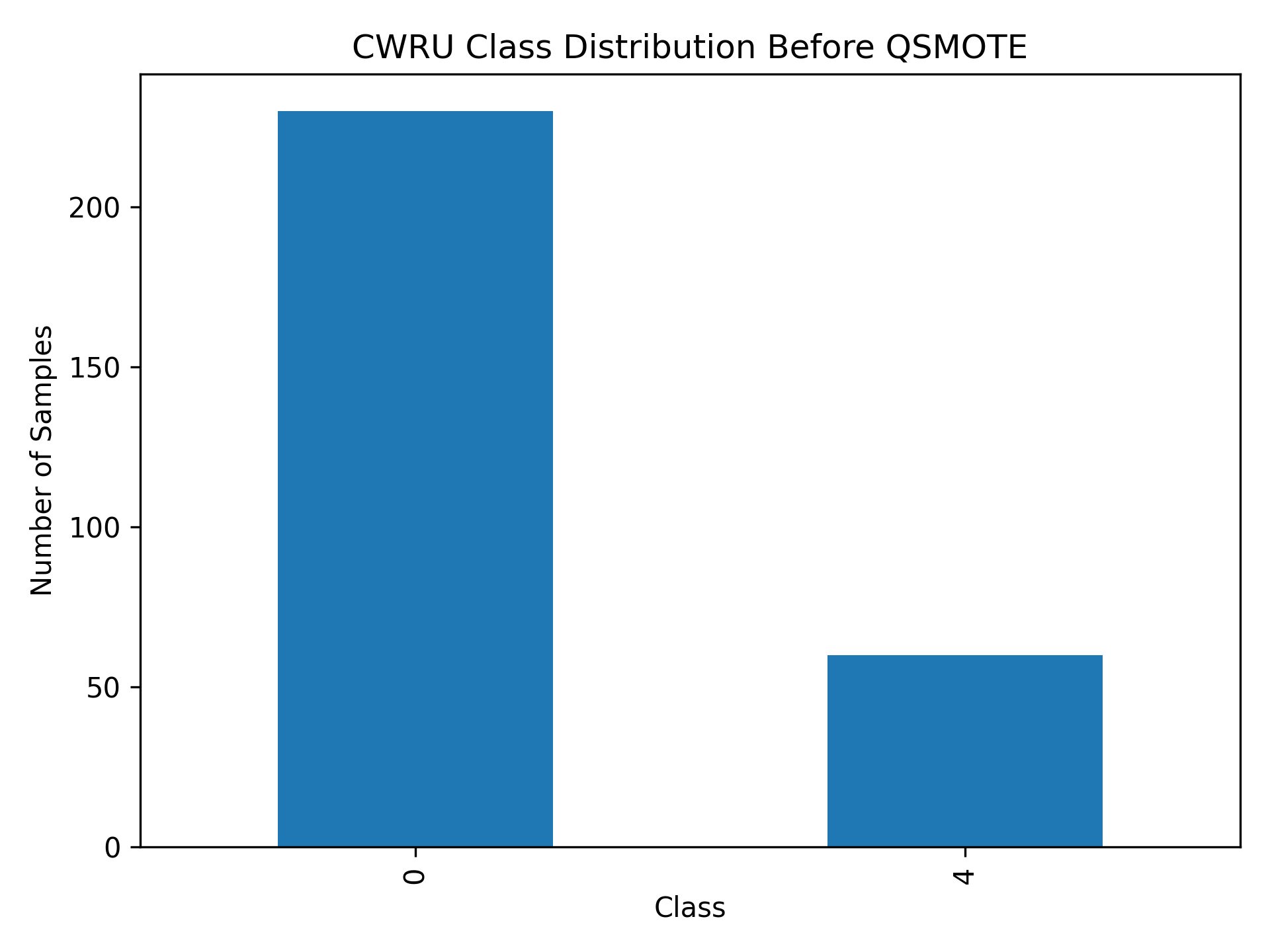}
        \caption{CWRUBD.}
        \label{fig:sup_cwrubd_before_qsmote}
    \end{subfigure}
    \hfill
    \begin{subfigure}[b]{0.235\textwidth}
        \centering
        \includegraphics[width=\linewidth]
        {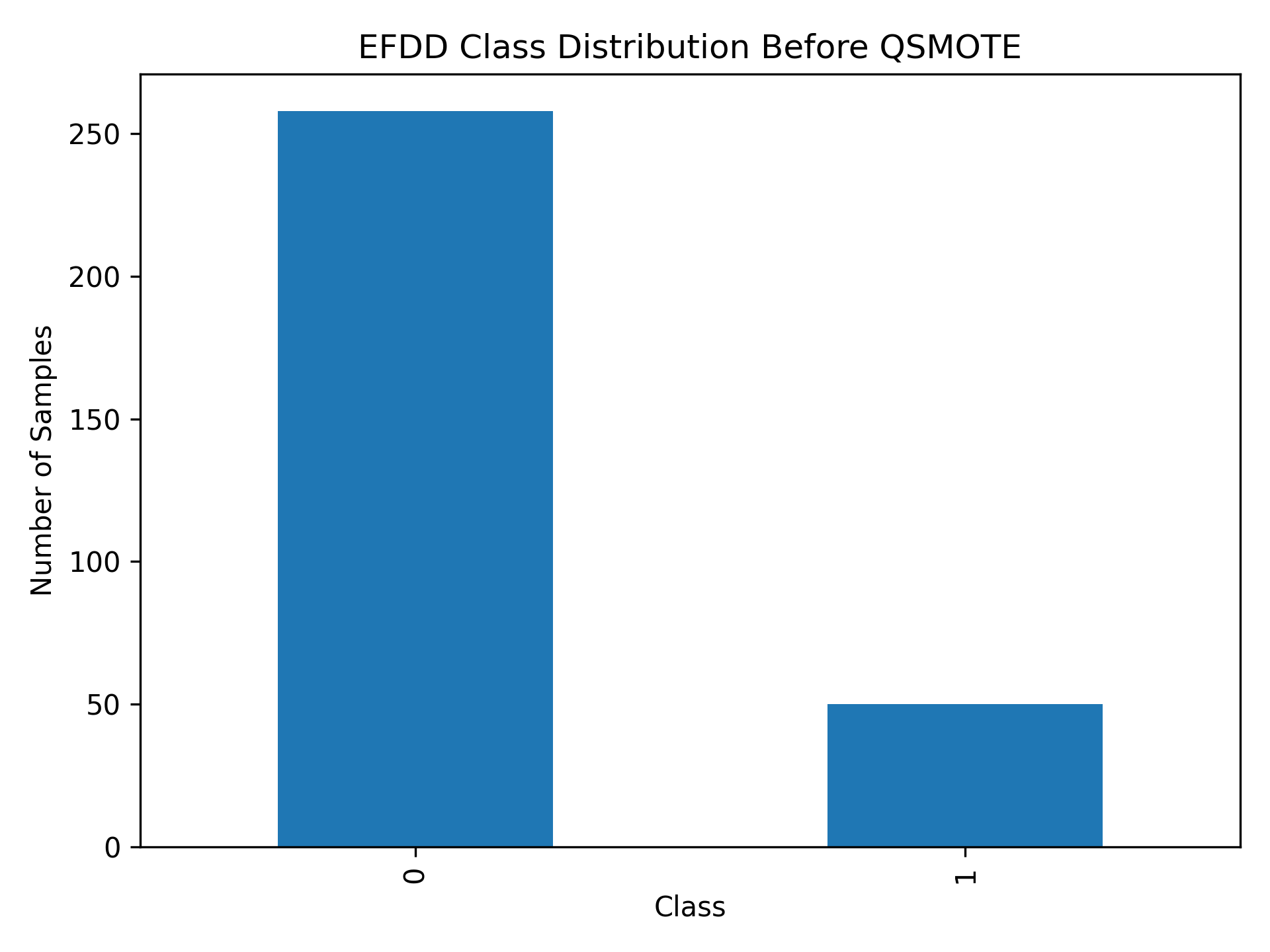}
        \caption{EFDD.}
        \label{fig:sup_efdd_before_qsmote}
    \end{subfigure}
    \hfill
    \begin{subfigure}[b]{0.235\textwidth}
        \centering
        \includegraphics[width=\linewidth]
        {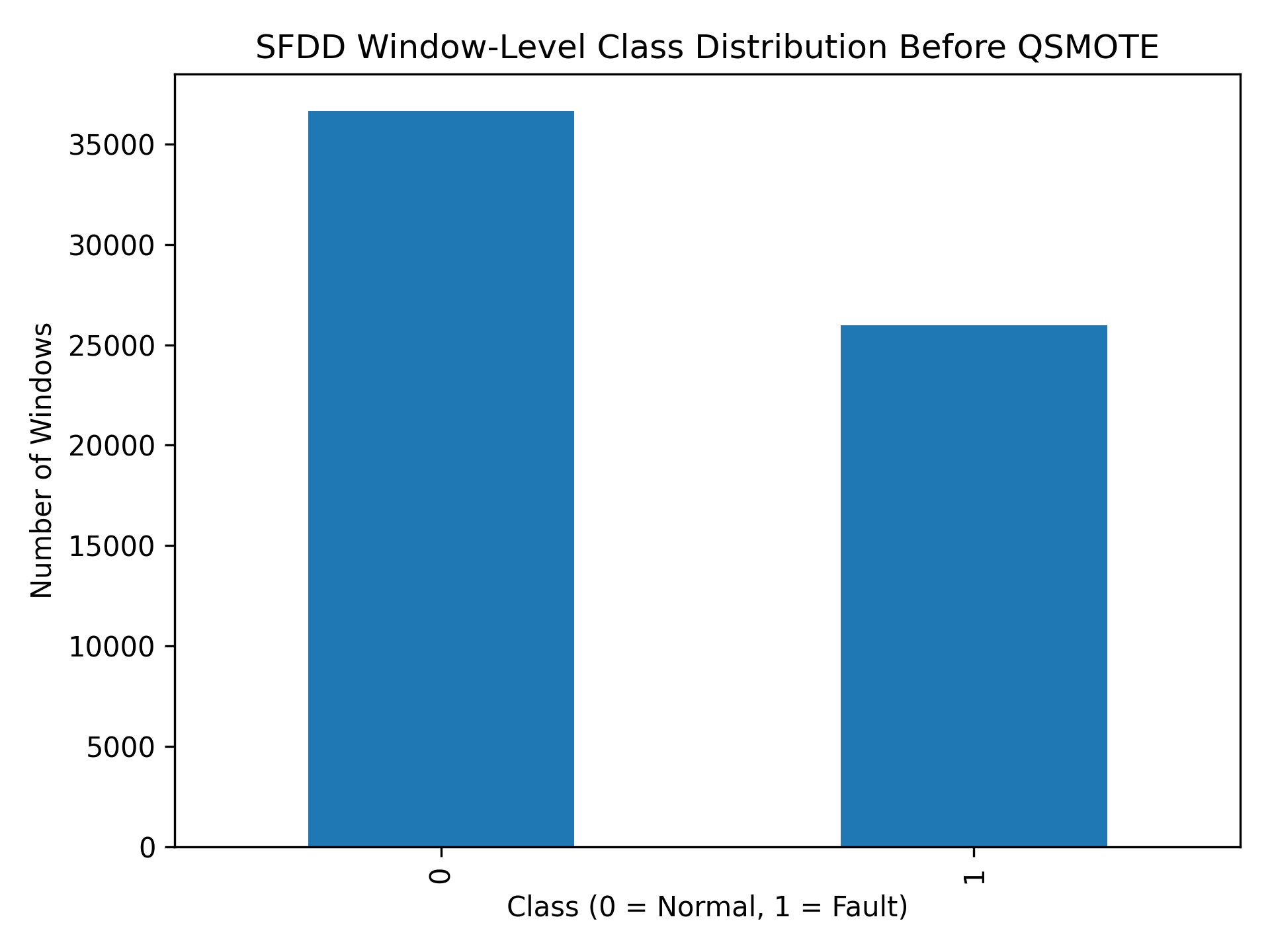}
        \caption{SFDD.}
        \label{fig:sup_sfdd_before_qsmote}
    \end{subfigure}

    \caption{Class distributions before QSMOTE for
    (a) SPID, (b) CWRUBD, (c) EFDD, and (d) SFDD.}
    \label{fig:sup_dataset_before_qsmote}
\end{figure*}
\iffalse
\begin{figure*}[!t]
    \centering

    \begin{subfigure}[b]{0.235\textwidth}
        \centering
        \includegraphics[width=\linewidth]
        {figures/SPID_QTN_RQNN_Confusion_Matrix.png}
        \caption{SPID.}
        \label{fig:sup_spid_cm}
    \end{subfigure}
    \hfill
    \begin{subfigure}[b]{0.235\textwidth}
        \centering
        \includegraphics[width=\linewidth]
        {figures/CWRU_QTN_RQNN_Confusion_Matrix.png}
        \caption{CWRUBD.}
        \label{fig:sup_cwru_cm}
    \end{subfigure}
    \hfill
    \begin{subfigure}[b]{0.235\textwidth}
        \centering
        \includegraphics[width=\linewidth]
        {figures/EFDD_QTN_RQNN_Confusion_Matrix.png}
        \caption{EFDD.}
        \label{fig:sup_efdd_cm}
    \end{subfigure}
    \hfill
    \begin{subfigure}[b]{0.235\textwidth}
        \centering
        \includegraphics[width=\linewidth]
        {figures/SFDD_QTN_RQNN_Confusion_Matrix.png}
        \caption{SFDD.}
        \label{fig:sup_sfdd_cm}
    \end{subfigure}

    \caption{Test-set confusion matrices of QTN-RQNN at the
    validation-selected decision thresholds for
    (a) SPID, (b) CWRUBD, (c) EFDD, and (d) SFDD.}
    \label{fig:sup_all_cm}
\end{figure*}
\fi

Algorithm~\ref{alg:qtn_rqnn} summarizes the complete training and
inference procedure of QTN-RQNN. The algorithm combines
training-only class balancing, sequential quantum feature encoding,
QTN-structured recurrent processing, quantum-memory propagation, and
hybrid quantum-classical optimization. The preprocessing transformations are estimated exclusively from the
training partition to prevent information leakage, and QSMOTE is
applied only to the training data. Each resulting input is divided
into $T$ sequential segments. At timestep $t$, the current two-feature
segment is encoded on the two input qubits, while the two memory
qubits contain the reduced state propagated from timestep $t-1$.
The combined input-memory state is processed by the QTN cell, whose
parameters $\boldsymbol{\theta}$ are shared across all recurrent
timesteps. After each application, the input subsystem is traced out,
yielding the updated memory state $\rho_{M,t}$. After the final timestep, local Pauli-$Z$ expectation values of the
two memory qubits form the readout vector $\mathbf{z}$. A classical
sigmoid layer maps this vector to the fault probability $\hat{p}$,
and the quantum and classical parameters are optimized using COBYLA
with class-weighted binary cross-entropy. The classification threshold
$\tau$ is selected exclusively from the validation partition and is
then fixed for test evaluation. Thus, the same trained QTN cell is
reused throughout the sequence, while recurrent information is carried
through the propagated quantum-memory state.

%\section*{Additional Experimental Results}\label{SecIV}

\subsection{Class Distributions Before QSMOTE}
\label{sup:qsmote_distribution}

Figure~\ref{fig:sup_dataset_before_qsmote} shows the class
distributions of the four fault-diagnosis datasets before
QSMOTE-based balancing. The datasets exhibit different degrees of
class imbalance, motivating the evaluation of QSMOTE within the
QTN-RQNN framework. QSMOTE is applied exclusively to the training
partition, while the validation and test partitions retain their
original class distributions.

\iffalse
\subsection{Confusion-Matrix Analysis}
\label{sup:confusion_matrices}

Figure~\ref{fig:sup_all_cm} presents the test-set confusion matrices
of QTN-RQNN at the validation-selected decision thresholds. On SPID,
the model correctly identifies 7 of the 10 fault samples while
retaining 19 true-negative predictions. For CWRUBD and EFDD, all
fault instances are detected, but this increased sensitivity is
accompanied by a substantial number of normal samples being classified
as faults. The effect is most pronounced for SFDD, where all 5,193
fault windows are detected but all 7,332 normal windows are also
assigned to the fault class. Therefore, the perfect recall observed
for CWRUBD, EFDD, and SFDD should be interpreted jointly with
precision and the threshold-independent ROC- and PR-AUC measures
reported in the main manuscript.
\fi

\normalsize
    % \begin{algorithm}[t]

    % \DontPrintSemicolon
    % \nonl\textbf{Require} $CS$, $QS$\;
    % \For{Voter $V_{i} \in \mathcal{V}$}{
    %     \textbf{Revelation}: \textbf{$\forall$} commitments $V_{i} \Rightarrow M_{j}$: Reveal $CS_{V_{i}}^{M_{j}}$
    % }
    % \For{Miner $M_{i} \in \mathcal{M}$}{
    %     {\textbf{Unswapping:} Perform the SWAPS from $CS_{V_{j}}^{M_{i}}$}\;
    %     {\textbf{Opening:} Miner M$_{i}$ begins restoring the $BC_{V_{j}}^{M_{i}}$ bits committed by Voter V$_{j}$}\;
    %     \textbf{Restore} {$BC_{V_{j}}^{M_{i}}$}\;
    %     {$BC_{V_{j}}^{M_{i}} \Rightarrow 00$: $R_X(0)$(CBM state) = Original state}\;
    %     {$BC_{V_{j}}^{M_{i}} \Rightarrow 01$: $R_X(\pi)$(CBM state) = Original state}\;
    %     {$BC_{V_{j}}^{M_{i}} \Rightarrow 10$: $R_X(\pi/2)$(IBM state) = Original state }\;
    %     {$BC_{V_{j}}^{M_{i}} \Rightarrow 11$: $R_X(3\pi/2)$(IBM state) = Original state }
    % }
    % \For{Voter $V_{i} \in \mathcal{V}$}{
    %     \textbf{Consensus}: Miners ($\mathcal{M}$) reach a consensus on the bit committed by V$_{i}$
    % }
    % \caption{Algorithm to perform ballot decommitment from voters ($\mathcal{V}$) to miners ($\mathcal{M}$). Here, CBM/IBM state refers to the state yielded from correct/incorrect basis measurement.} \label{algo:algo2} 
    % \end{algorithm}

\ifCLASSOPTIONcaptionsoff
  \newpage
\fi

\vfill


\end{document}